\documentclass[twocolumn,english,aps,pre,showpacs,superscriptaddress]{revtex4-2}

\usepackage{graphicx}
\usepackage{dcolumn}
\usepackage{bm}
\usepackage{hyperref}

\usepackage[T1]{fontenc}
\usepackage[latin9]{inputenc}
\usepackage{mathrsfs}
\usepackage{amssymb}
\usepackage{amsmath}
\usepackage{wasysym}
\usepackage{esint}
\usepackage{babel}
\usepackage{xcolor}
\usepackage{physics}

\begin{document}

\title{Nonclassicality from Incomplete Fock-State Statistics}

\author{Martin \v{Z}onda}

\address{Department of Condensed Matter Physics, Faculty of Mathematics and Physics, Charles University, Ke Karlovu 5, Praha 2 CZ-121 16, Czech Republic}

\author{Pavel Bal\'a\v{z}}

\address{FZU -- Institute of Physics of the Czech Academy of Sciences, Na Slovance 1999/2, 182 21 Prague 8, Czech Republic}

\author{Luk\'a\v s Lachman}
\address{Department of Optics, Palack\'y University, 17. listopadu 1192/12, 779 00 Olomouc, Czech Republic}

\author{Darren W. Moore} 
\address{Department of Optics, Palack\'y University, 17. listopadu 1192/12, 779 00 Olomouc, Czech Republic}

\author{Tom\'a\v s Novotn\' y}
\address{Department of Condensed Matter Physics, Faculty of Mathematics and Physics, Charles University, Ke Karlovu 5, Praha 2 CZ-121 16, Czech Republic}

\author{Radim Filip}
\address{Department of Optics, Palack\'y University, 17. listopadu 1192/12, 779 00 Olomouc, Czech Republic}
\date{\today}

\begin{abstract}

Nonclassicality in bosonic systems can be certified directly from occupation probabilities without full state reconstruction. To complete a library of such criteria we introduce a numerical framework that constructs optimal linear nonclassicality witnesses from arbitrary subsets of measured occupation probabilities by solving a semi-infinite linear program. The method contains known analytical criteria as special cases while generally detecting substantially larger nonclassical regions. We provide an open-source implementation of the numerical optimization together with the analytical criteria discussed here, enabling direct application of the framework to measured occupation statistics across bosonic platforms. As a case study, we apply it to the resonant Jaynes--Cummings model with thermal states and reveal nonclassicality at short interaction times and close to thermal balance, where detection of nonclassicality is otherwise challenging. 
\end{abstract}
\maketitle


{\it Introduction }---~The identification of nonclassical states of harmonic oscillators is a central problem in bosonic systems. Since the seminal works of Glauber and Sudarshan~\cite{Glauber1963coherent,Sudarshan1963equivalence}, nonclassicality has been associated with the impossibility of representing a quantum state as a statistical mixture of coherent states. Such nonclassical states constitute an essential requirement for quantum technologies across a wide range of bosonic platforms. At the same time, reliable certification of nonclassicality remains challenging because experimentally accessible information is often influenced by thermal noise and detection is limited. Consequently, many experimental platforms and regimes remain unexplored.

Many nonclassicality criteria require either full state reconstruction or the accurate measurement of high-order moments of bosonic variables. In a broad range of experimental platforms, including quantum optics~\cite{sperling_detector-independent_2017,Lita2022Development,Sonoyama2022Analysis,Cheng2023Pixel,eaton_resolution_2023,Hlousek2024High,Ding2025Photon,Endo2025Optically,Schapeler2026macroscopic,Schapeler2026Practical}, cavity~\cite{guerlin_progressive_2007,deleglise_reconstruction_2008,brune_process_2008,sarlette_stabilization_2011} and circuit quantum electrodynamics~\cite{Schuster2007Resolving,Afzelius2010Photon,Harder2017Single,kono_nonclassical_2017,Lee2020Macroscopic,Owens2022Chiral}, trapped ions~\cite{Ding2017Quantum,Ding2017Cross,Wolf2019Motional,Chen2023Scalable}, and optomechanical systems~\cite{Shomroni2018All,Sivak2018Kerr,Besse2020Single}, modern boson-number-resolving detection schemes offer the most directly accessible information given by, at least incomplete, occupation probabilities of the Fock states. 
A particularly influential family of such criteria was introduced by Klyshko~\cite{klyshko1996observable}, who derived inequalities between neighboring occupation probabilities that must be satisfied by every classical state. Their violation provides a simple and experimentally accessible witness of nonclassicality.

More recently, these ideas have been generalized in several directions, including criteria in low-dimensional probability spaces~\cite{filip_hierarchy_2013}, criteria based on photon-number distributions~\cite{Perina2020Non} or quantum state distances~\cite{ares_distance-based_2022}, observable witnesses for multiplexed detection systems~\cite{krishnaswamy_experimental_2024,kovtoniuk_tight_2024,fiurasek_tight_2024,fiurasek_fundamental_2025,krishnaswamy2026observable,Kovtoniuk2026nonclassical}, and majorization-based constructions involving higher-order occupation probabilities~\cite{medranda_sub-poissonian_2015,innocenti_nonclassicality_2022,Kovtoniuk2026nonclassical}. These analytical developments demonstrated that finite sets of occupation probabilities contain sufficient information to certify nonclassicality directly and stimulated growing interest in Fock-probability-based approaches to nonclassicality detection. Related recent efforts have also explored data-driven and machine-learning-assisted strategies for extracting nonclassical features from experimentally accessible counting statistics~\cite{machulka2024revealing,jung2026learning}.

Despite this progress, obtaining a complete set of analytical witnesses, i.e., witnesses defined by closed analytical expressions, rapidly becomes difficult as the number of available probabilities increases. Consequently, existing analytical criteria typically exploit only a fraction of the information contained in measured occupation statistics. Simultaneously, it is challenging to generally predict where nonclassical aspects could emerge in occupation probabilities and remain robust in the nonlinear, noisy dynamics of bosonic systems.

In this work, we overcome these limitations by formulating nonclassicality
certification from arbitrary subsets of occupation probabilities as a
semi-infinite linear program. We exploit
the Poisson-mixture structure of classical Fock-state statistics to directly
test whether measured occupation probabilities are compatible with any
mixture of coherent states. The method naturally reproduces the known
analytical criteria and, typically, identifies substantially larger
nonclassical regions. Moreover, it allows for direct identification of where
nonclassicality appears in the Fock basis.

To facilitate its use, we provide a fast open-source Python implementation of the numerical algorithm, together with code that automatically generates all analytical criteria considered in this work~\cite{code_github}.~\footnote{Note, that linear-programming approaches have previously
been developed for generalized noncontextuality~\cite{Gitton2022solvablecriterion,
Selby2024linear,rossi2026theory}, which addresses a distinct notion of
classicality for prepare-and-measure statistics.}

{\it Nonclassicality criteria }---~
A classical state always corresponds to a statistical mixture of coherent states~\cite{Sudarshan1963equivalence}. Consequently, a necessary condition for classicality is that the vector of measured Fock-state probabilities ${P_k}$ can be expressed as a convex combination of Poisson distributions, $p_k(\mu)=e^{-\mu}\mu^k/k!$, which are the photon-number distributions of coherent states. Any probability vector that violates the constraints imposed by this classical set therefore certifies nonclassicality. 

This observation immediately leads to a general linear witness. For a finite set of occupation indices $\mathcal{I}$, we define the measured-probability vector
$\bm{P}=(P_i)_{i\in\mathcal{I}}$, the corresponding Poisson vector
$\bm{p}(\mu)=(p_i(\mu))_{i\in\mathcal{I}}$, and the weight vector
$\bm{\lambda}=(\lambda_i)_{i\in\mathcal{I}}$, with 
$\lambda_i\in\mathbb{R}$. Nonclassicality is certified whenever
\begin{equation}
{\cal W}(\bm{\lambda};\bm{P})
=
\bm{\lambda}\cdot\bm{P}
-
\sup_{\mu\geq0}
\left(\bm{\lambda}\cdot\bm{p}(\mu)\right)
>0,
\label{eq:gcrb}
\end{equation}
where $\cdot$ denotes the dot product. It is sufficient that this inequality
holds for at least one choice of $\bm{\lambda}$.

{\it Numerical criteria}---~
We now turn to the central result of this work: the numerical construction of optimal linear nonclassicality witnesses within an arbitrary set of available occupation probabilities.
Casting the optimization of the witness~\eqref{eq:gcrb} as a convex minimization problem leads to a generally non-smooth objective. The non-smoothness arises because the point determining the supremum can change discontinuously as $\boldsymbol\lambda$ varies.
In practice, we found
standard numerical optimization methods to be unreliable.

To avoid the direct optimization of the non-smooth supremum, we introduce an
auxiliary variable $t$ that bounds the classical value of the witness,
\begin{equation}
t\geq\bm{\lambda}\cdot\bm{p}(\mu),
\qquad \forall\,\mu\geq0.
\label{eq:num_impl_constraint}
\end{equation}
For a fixed $\bm{\lambda}$, minimizing $t$ therefore drives it toward the supremum of $\bm{\lambda}\cdot\bm{p}(\mu)$ over all $\mu \geq 0$. The original problem is thus exactly equivalent to
\begin{equation}
\min_{\bm{\lambda},t}
\left(t-\bm{\lambda}\cdot\bm{P}\right),
\label{eq:num_impl_lp}
\end{equation}
subject to Eq.~\eqref{eq:num_impl_constraint}, $t\geq0$, and
$-1\leq\lambda_j\leq1$. The condition $t\geq0$ accounts for the limit
$\bm{p}(\mu)\rightarrow0$ as $\mu\rightarrow\infty$. The bounds on
$\lambda_j$ only fix the normalization of the witness and do not restrict
the detectable states: since Eq.~\eqref{eq:gcrb} is homogeneous in
$\bm{\lambda}$, any witness can be rescaled to satisfy
$\max_j|\lambda_j|=1$. Importantly, for every
fixed $\mu$ Eq.~\eqref{eq:num_impl_constraint} is a linear inequality in
$(\bm{\lambda},t)$: each Poisson distribution therefore provides one linear
constraint on the possible witness. The objective~\eqref{eq:num_impl_lp} is
linear as well. The only complication is that $\mu$ is continuous, giving
infinitely many such constraints.  The resulting problem is therefore a semi-infinite linear program~\cite{Vanderbei2001linear}.

We solve this problem iteratively using an exchange (cutting-plane)
procedure~\cite{Hettich1993,Vanderbei2001linear}. We first retain only a
finite set of Poisson parameters $\mathcal M=\{\mu_1,\mu_2,\ldots\}$ and
solve the resulting ordinary linear program. This gives a candidate
$(\bm{\lambda},t)$ that satisfies the classical bound at all points currently
included in $\mathcal M$, but may still violate it at some omitted value of
$\mu$. We therefore test the candidate against the entire Poisson family by
calculating
\begin{equation}
w=\sup_{\mu\geq0}\bm{\lambda}\cdot\bm{p}(\mu).
\label{eq:num_impl_worst_case}
\end{equation}
If $w\leq t+\delta$, all omitted constraints are satisfied within the
numerical tolerance $\delta$, and the solution is accepted. 
Otherwise, the supremum is attained at a finite point $\mu^\ast$ which identifies the most strongly violated constraint.
We add $\mu^\ast$ to $\mathcal M$, solve the enlarged linear program, and
repeat the procedure. In this way the relevant parts of the continuous
classical boundary are generated only as they are needed, rather than being
approximated by a predetermined discretization of $\mu$.

The remaining maximization in Eq.~\eqref{eq:num_impl_worst_case} is only
one-dimensional and can be performed efficiently. For fixed
$\bm{\lambda}$,
$
g(\mu)=\bm{\lambda}\cdot\bm{p}(\mu)
=e^{-\mu}\sum_k\lambda_k \mu^k/k!,
$ where coefficients corresponding to unavailable $P_k$ are set to zero.
Its stationary points are roots of a finite polynomial obtained from
$dg/d\mu=0$. Evaluating $g(\mu)$ at all non-negative real stationary
points and at $\mu=0$, together with the limiting value zero as $\mu\rightarrow\infty$, yields the supremum in Eq.~\eqref{eq:num_impl_worst_case}.  Thus, the original non-smooth
optimization is replaced by a sequence of standard finite linear programs
and one-dimensional maximizations. In practice, only a small number of
constraints needs to be added before convergence. Further mathematical and
implementation~\cite{code_github} details are given in the SM~\cite{SM}.

{\it Analytical criteria }---~We compare the numerical approach with three families of analytical
nonclassicality criteria. Some of these witnesses are presented here in closed analytical form, while for others we provide a Python implementation that generates them automatically~\cite{code_github}. Additional criteria, technical details, and numerical tests are provided in the Supplemental Material (SM)~\cite{SM}.

When only two Fock-state probabilities, $P_0$ and $P_n$, are available, one can employ the analytical conditions
\begin{equation}
\begin{split}
P_1 &> -P_0 \ln P_0,\\
P_{n>1} &>
\begin{cases}
\dfrac{P_0}{n!}(-\ln P_0)^n, & 0<P_0\le P_0^c,\\[1.2ex]
C_n\left(1-P_0\right), & P_0^c < P_0\le 1,
\end{cases}
\end{split}
\label{eq:0n}
\end{equation}
where $P_0^c=e^{-x}$, $x$ is the nonzero real solution of $x=n(1-e^{-x})$, i.e., $x=n+\mathrm{LambertW}(-n e^{-n})$, and $C_n=x^{n-1}P_0^c/(n-1)!$. This construction is equivalent to that of Ref.~\cite{filip_hierarchy_2013}, now derived in a closed analytical form. 

For three consecutive probabilities, one obtains the standard Klyshko criteria~\cite{klyshko1996observable} and their complement~\cite{innocenti_nonclassicality_2022}
\begin{equation}
\begin{split}
(k+1)P_{k-1}P_{k+1} - k P_k^2 &< 0,\qquad k>1,\\
P_0 + \frac{P_1^2}{2P_2}\left[\exp\left(\frac{2P_2}{P_1}\right)-1\right] &> 1.
\end{split}
\label{eq:Klyshko}
\end{equation}

More recently, generalized conditions involving higher Fock-state probabilities were derived in Ref.~\cite{innocenti_nonclassicality_2022},
\begin{equation}
N P_N\!>\!P_{N-1}\ln\!\left[
\sum_{k=0}^{N-1}\!\frac{N^kP_N^k}{k!P_{N-1}^k}\!+\!
\frac{N^NP_N^{N-1}}{N!P_{N-1}^N}\!\left(\!1\!-\!\sum_{k=0}^{N-2}P_k\!\right)
\!\right].
\label{eq:KlyshkoN}
\end{equation}
Whenever applicable, we employ any of above witnesses and collectively refer to them as ${\cal W}^{\cal K}$ witness family.

The second large class of analytical criteria considered here is based on generalized Klyshko-type inequalities obtained via a majorization approach~\cite{innocenti_nonclassicality_2022}. One considers two integer tuples $T^{A}$ and $T^{B}$ satisfying $T^{A}\succcurlyeq T^{B}$. From such a pair one obtains
\begin{equation}
\prod_{i=1}^s T^{B}_i!\, P_{T^{B}_i}
\;\leq\;
\prod_{i=1}^s T^{A}_i!\, P_{T^{A}_i},
\label{eq:GKM}
\end{equation}
which must hold for all classical states. Violations of Eq.~\eqref{eq:GKM} therefore define a family of nonclassicality witnesses, denoted by ${\cal W}^{\cal M}$. 

The final class of criteria used here has recently been formulated in terms of the counting matrix~\cite{krishnaswamy2026observable}. For a set of either integer or half-integer indices $\mathcal{I}$, chosen such that all $k+l$ correspond to available occupation probabilities, one constructs the largest possible counting matrices with elements
\begin{equation}
C^{(\mathcal{I})}_{kl}=(k+l)!\,P_{k+l},
\qquad k,l\in\mathcal{I}.
\label{eq:counting_matrix}
\end{equation}
For every classical state, the counting matrix $C^{(\mathcal{I})}$ must be positive semidefinite. Thus, a negative eigenvalue of any admissible counting matrix certifies nonclassicality. We denote the resulting family of witnesses by ${\cal W}^{\cal C}$. 
Note that this criterion requires at least three occupation probabilities $P_i$, $P_j$, $P_k$ whose indices $i,j,k$ form an arithmetic progression. Further details, including the construction with half-integer indices, are given in Ref.~\cite{krishnaswamy2026observable} and SM~\cite{SM}.
All the above analytical criteria for an arbitrary set of occupation probabilities ${P_k}$ are implemented in the provided open-source code~\cite{code_github}.

{\it Numerical vs analytical boundaries }---~
\begin{figure*}[ht!]
	\includegraphics[width=0.95\textwidth]{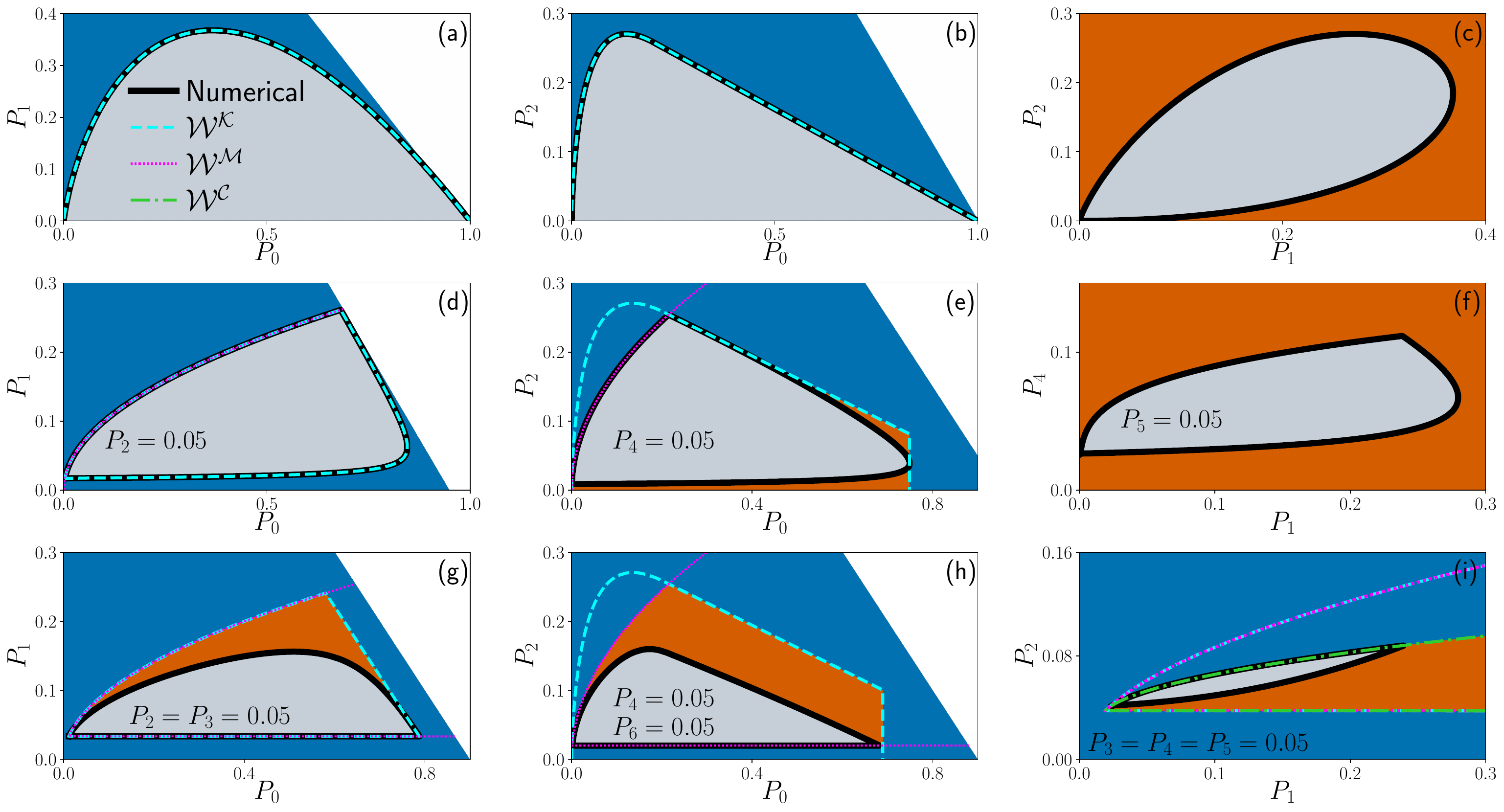}\caption{
Comparison of nonclassical regions identified by the numerical criterion and analytical witnesses in reduced spaces spanned by selected sets of probabilities $\{P_k\}$. Blue regions correspond to states identified as nonclassical by both approaches, while orange regions indicate nonclassical states detected only by the numerical optimization. White regions denote nonphysical probability combinations with $\sum_k P_k>1$. Solid black curves show the nonclassicality boundaries obtained numerically. Cyan dashed curves correspond to the witnesses ${\cal W}^{\cal K}$, magenta dotted curves to the majorization witnesses ${\cal W}^{\cal M}$, and green dash-dotted curves to the counting-matrix witnesses ${\cal W}^{\cal C}$. The latter are shown only in panel (i), since in all other panels where available their boundaries coincide with those obtained from ${\cal W}^{\cal M}$. 
\label{fig:NoModel}}
\end{figure*}
To assess the numerical criterion independently of a particular physical
model, we compare it directly with the analytical witnesses in reduced
probability spaces.
Figure~\ref{fig:NoModel} shows representative examples of nonclassicality boundaries obtained from numerical optimization (black curves) and from the available analytical criteria. The analytical results are constructed from all applicable Klyshko-type witnesses ${\cal W}^{\cal K}$ (cyan dashed curves), majorization witnesses ${\cal W}^{\cal M}$ (magenta dotted curves), and counting matrix witnesses ${\cal W}^{\cal C}$ (green dash-dotted curves). Blue regions correspond to states identified as nonclassical by both numerical and analytical criteria, whereas orange regions represent additional nonclassical states detected exclusively by the numerical approach. Panels (a--c), (d--f), (g,h), and (i) use two, three, four, and five probabilities, respectively.  Note that the pictures are zoomed in to the interesting regions in the vicinity of the numerical boundary and white regions mark non-physical combinations of probabilities.

The selected examples illustrate several important features. Panels (a), (b), and (d) show cases in which the available Klyshko witnesses ${\cal W}^{\cal K}$ reproduce exactly the same boundary as obtained from the numerical optimization. In fact, we have not found any counterexample in the $(P_0,P_n)$ or $(P_0,P_1,P_2)$ probability spaces where the numerical boundary differs from the analytical one. Additional examples and tests are presented in the SM~\cite{SM}.

However, this agreement does not generally persist when more probabilities are available. As illustrated in panel (g), when four consecutive probabilities are used, the numerical criterion identifies a strictly larger nonclassical region. We also observe that, whenever all probabilities up to a given order $N$ are available, the majorization witnesses ${\cal W}^{\cal M}$ provide no additional coverage beyond that already obtained from ${\cal W}^{\cal K}$. 

The situation is markedly different for nonconsecutive occupation probabilities. Panels (e), (h) show that the majorization witnesses ${\cal W}^{\cal M}$ can substantially enlarge the nonclassical region detected by ${\cal W}^{\cal K}$. This demonstrates that majorization-based criteria become particularly useful when only selected higher-order probabilities are accessible. On the other hand, with at least five indices that form an arithmetic progression (for example consecutive probabilities), i.e., the minimal set required to construct a $3\times3$ counting matrix, the counting-matrix witnesses ${\cal W}^{\cal C}$ can become stronger than both ${\cal W}^{\cal M}$ and ${\cal W}^{\cal K}$, as illustrated in panel (i).
Nevertheless, even in these cases the numerical criteria cover a larger probability space than combined analytical ones. 

Finally, panels (c) and (f) highlight the main advantage of the numerical approach. In these examples, neither Klyshko nor majorization or counting matrix witnesses can be constructed from the available probabilities. Nevertheless, the numerical optimization reveals that a large fraction of the $(P_1,P_2)$ and $(P_1,P_4,P_5=0.05)$ spaces corresponds to nonclassical states. This demonstrates that the numerical criteria are advantageous even in cases where only a small subset of occupation probabilities is known.

It is important to emphasize that the numerical criterion is the most general witness considered here. Every state identified as nonclassical by an analytical witness is also detected by the numerical optimization, while the converse is not true, as demonstrated by the orange regions in Fig.~\ref{fig:NoModel}. These regions correspond to nonclassical states for which a valid linear witness exists but is not captured by the available analytical criteria. Having established this advantage in a model-independent setting, we now examine its consequences for a paradigmatic bosonic noisy system.

{\it Jaynes--Cummings model }---~ 
 We consider the resonant Jaynes--Cummings (JC) model, a physically and technologically important model describing the interaction between a two-level system and a bosonic mode~\cite{DeBernardis2024cavity,Larson2024jaynes}. 
This paradigmatic light--matter interaction model is known to deterministically generate nonclassical bosonic states even when both subsystems are initially prepared in (distinct) thermal states~\cite{marek2016deterministic,Slodicka2016deterministic}, and the nonclassical aspects of these states have recently attracted renewed attention~\cite{Yang2025Hot}.
Its Hamiltonian is
$
H =
\hbar\omega\sigma_z/2
+\hbar\omega a^\dagger a
+\hbar g \left(a^\dagger\sigma_- + a\sigma_+\right),
$
where $\omega$ is the transition frequency of the two-level system equal to the oscillator frequency (resonant condition),
$g$ denotes the coupling strength,
$a^\dagger$ and $a$ are the bosonic creation and annihilation operators,
and $\sigma_z$, $\sigma_+$, and $\sigma_-$ are the usual Pauli and ladder operators of the two-level system. We assume that, initially, each subsystem is locally in thermal equilibrium, characterized by different temperatures $T_o$ and $T_q$ for the oscillator and the two-level system (qubit), respectively. This scenario was analyzed in detail in Ref.~\cite{marek2016deterministic} using analytical criteria. 

The initial state of the two-level system is
$
\rho_q
=
p_e |e\rangle\langle e|
+
(1-p_e)|g\rangle\langle g|,
$
with excited-state population given by Boltzmann distribution 
$
p_e=
\exp(-\frac{\hbar\omega}{k_B T_q})/
     [1+\exp(-\frac{\hbar\omega}{k_B T_q})]
$.
The oscillator is prepared in a thermal state
$
\rho_o
=
\sum_{n=0}^{\infty}
\bar n^n/(1+\bar n)^{n+1}
|n\rangle\langle n|,
\label{eq:rhoth}
$
where the mean occupation number
$
\bar n=
1/[
\exp(\hbar\omega/(k_B T_o))-1]
$ is determined by the oscillator temperature. Tracing over the two-level system, the oscillator state evolves as
$
\rho_o(\tau)
=
\sum_{n=0}^{\infty}
P_n(\tau)
|n\rangle\langle n|,
$
where $\tau=gt$ and $P_n(\tau)$ are given explicitly in Eq.~(129) in SM~\cite{SM}. Note that when both subsystems are initially at the same temperature, corresponding to the condition
$
\bar n=\bar{n}_e\equiv p_e/(1-2p_e),
$ the global thermal equilibrium state is recovered and no time evolution occurs. Here we focus only on a typical case where all first $N_P$ occupation probabilities are used to categorize a state. 

For clarity, the initial comparison is restricted to the Klyshko family ${\cal W}^{\cal K}$ as majorization family ${\cal W}^{\cal M}$ does not provide additional coverage for this case. For large $N_P$, the counting-matrix criteria reproduce the numerically identified boundary much more closely than ${\cal W}^{\cal K}$, and their
convergence is analyzed separately in connection with the thermal-balance
regime.

\begin{figure}[ht!]
    \includegraphics[width=1.0\columnwidth]{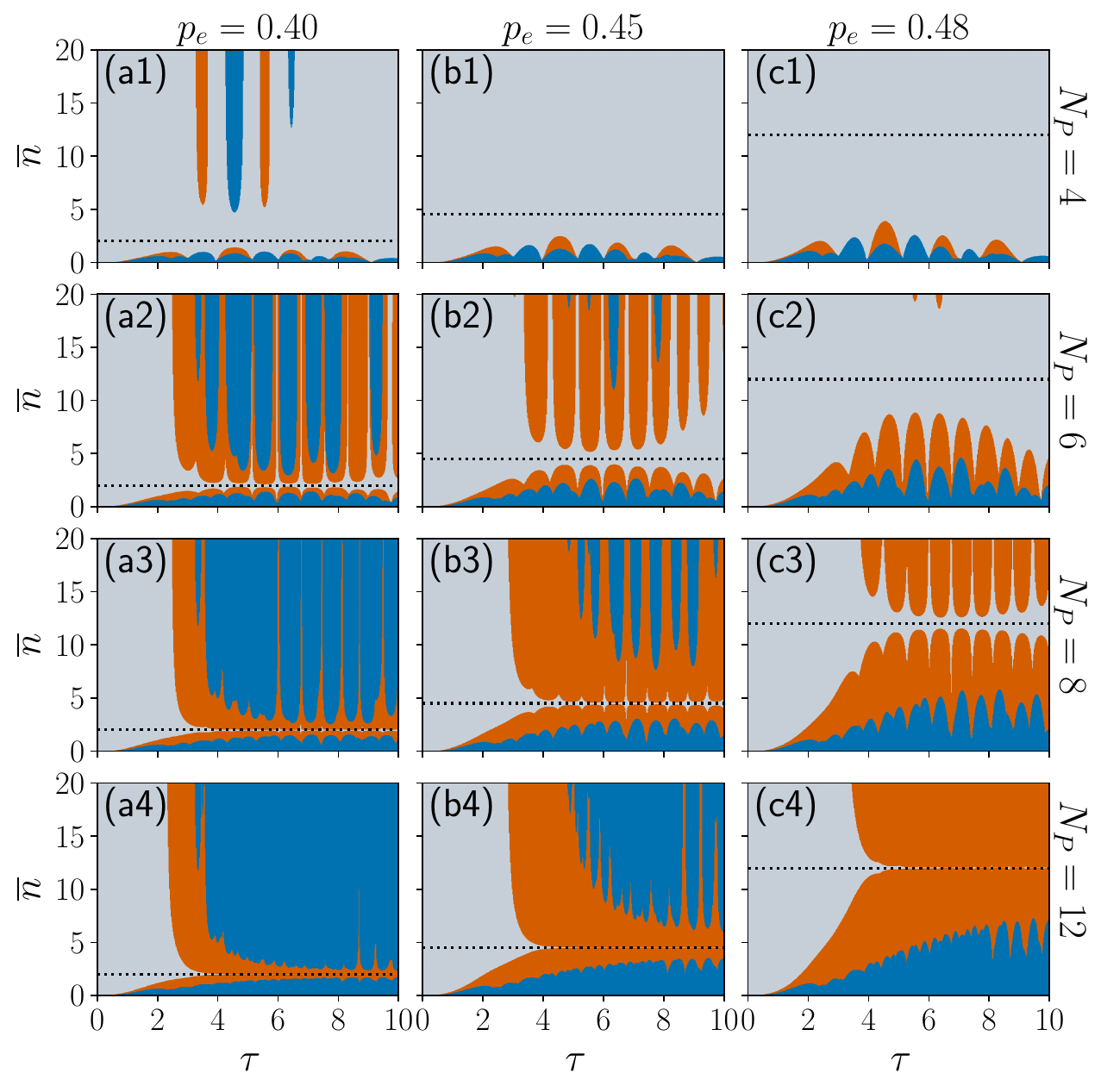}
    \caption{
    Comparison of nonclassical regions obtained using analytical ${\cal W}^{\cal K}$ and numerical criteria for the Jaynes--Cummings model for the initial excited-state population $p_e=0.4$, $0.45$, $0.48$, respectively $k_BT_q/\hbar\omega=2.47$, $4.98$,  $12.5$,  and different numbers of probabilities $N_P$ (see right vertical axis) used in the criteria. The dark blue region represents nonclassicality detected by the analytical criterion, while the orange regions correspond to additional nonclassicality detected only by the numerical criterion. The horizontal dotted line indicates the thermal-equilibrium condition given by $\bar n=\bar{n}_e\equiv p_e/(1-2p_e)$=$2$,$4.5$, $12$.
    }
    \label{fig:jc}
\end{figure}

Figure~\ref{fig:jc} shows the nonclassical regions identified by ${\cal W}^{\cal K}$ and by the numerical optimization for several values of $p_e$ and increasing numbers of available occupation probabilities $N_P$. As before, dark blue regions correspond to states identified as nonclassical by both approaches, while orange regions indicate additional nonclassical states detected only by the numerical criterion. Since the numerical and analytical boundaries coincide for $N_P\le 3$, only the results for $N_P\ge4$ are shown.

Two trends are immediately apparent in areas challenging for nonclassicality detection. First, the numerical criterion extends the detected nonclassical region toward shorter interaction times $\tau$ and larger initial $\bar{n}$, i.e., higher oscillator temperatures. 
Both are experimentally favorable, since shorter interaction times reduce exposure to
decoherence and higher $T_o$ requires less oscillator cooling.

Second, the greatest improvement occurs in the vicinity of the thermal-balance condition $
\bar n=\bar{n}_e$. As the number of available probabilities increases, the region which is not classified as nonclassical around the thermal balance shrinks substantially, revealing nonclassical states that remain undetected by the analytical witnesses in Eqs.~\eqref{eq:0n}, \eqref{eq:Klyshko}, and \eqref{eq:KlyshkoN}.

The advantage of the numerical criterion becomes increasingly pronounced as $p_e$ approaches $1/2$, corresponding to higher temperatures and larger equilibrium occupations $\bar n_e$. Although the state remains strictly classical at the equilibrium point $\bar n=\bar n_e$ (dashed horizontal line), the numerical analysis reveals sizable nonclassical regions in its immediate vicinity. Importantly, this appears to be a novel regime that is suitable for immediate experimental exploration.

\begin{figure}[!t]
\includegraphics[width=0.95\columnwidth]{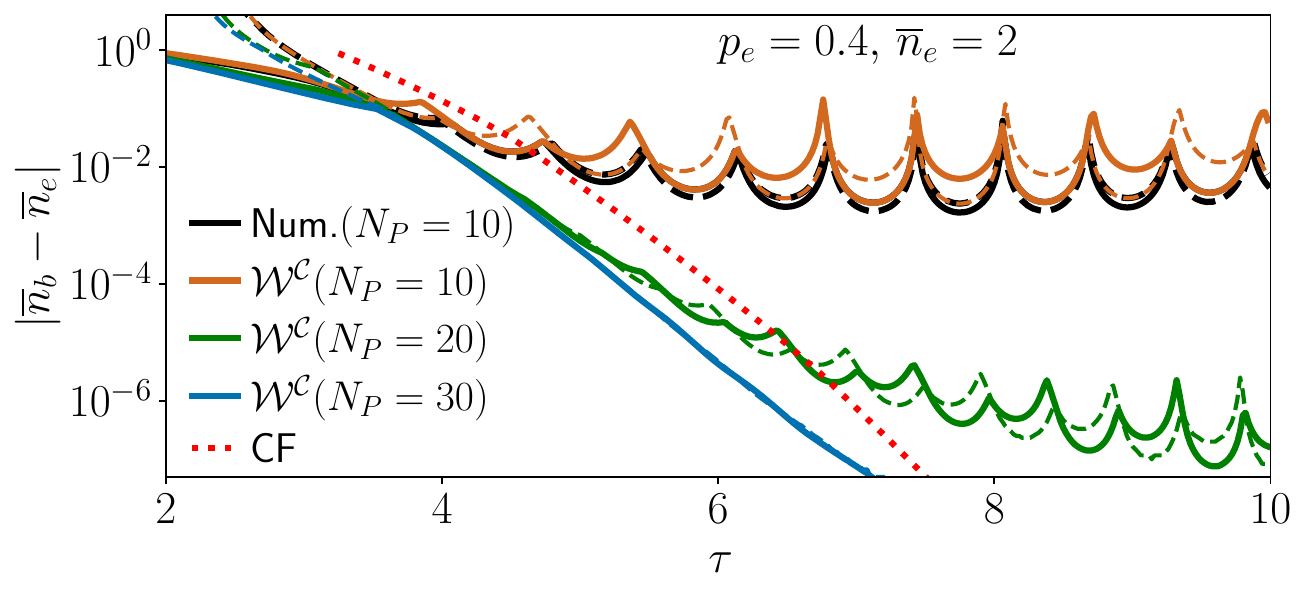}
    \caption{Distance of the nonclassicality boundaries from the thermal-balance condition $\bar n_{e}=2$ for $p_e=0.40$ [see Fig.~\ref{fig:jc}(a4)]. Solid and dashed lines correspond to the boundaries below and above $\bar n_{ e}$, respectively. Results are shown for the numerical criterion, the counting matrix criterion ${\cal W}^{\cal C}$ for different $N_P$, and the criterion based on the characteristic function (CF).
    }
    \label{fig:near_tb}
\end{figure}

Resolving nonclassicality close to thermal balance requires access to a larger number of occupation probabilities. 
Figure~\ref{fig:near_tb} examines this region in more detail by showing the distance of the upper (dashed lines) and lower (solid) nonclassicality boundaries from $\bar n_e=2$ and $p_e=0.40$. Even for large $N_P=10$, i.e., ten consecutive probabilities,  the numerical criterion (black lines) is closer to thermal balance than the counting criterion ${\cal W}^{\cal C}$ (brown lines), which is the best of the analytical criteria available here. Nevertheless, ${\cal W}^{\cal C}$ at sufficiently large $N_P$, as well as 
an independent check of nonclassicality using characteristic function criteria~\cite{Vogel2000nonclassical,Richter2002nonclassicality}, confirm the behavior revealed already by the numerical analysis: the classical region around thermal balance narrows exponentially with interaction time. Consequently, any finite thermal imbalance in principle leads to detectable nonclassicality, with the required interaction time increasing only logarithmically as the imbalance decreases. In addition, although not discussed here in detail, we have also checked that this effect is stable under small pure loss on the oscillator.

{\it Summary }---~We have introduced a numerical framework for certifying nonclassicality from arbitrary finite sets of occupation probabilities by optimizing linear witnesses through semi-infinite linear programming. For all cases considered here, the numerical optimization detects every nonclassical state identified by the analytical criteria and, typically, extends the detected region beyond them. Using the Jaynes--Cummings model applied to thermal states as a testing example, we showed that the numerical approach reveals nonclassicality regimes at short interaction times and close to thermal balance. We provide open-source code~\cite{code_github} implementing the numerical method together with all analytical criteria discussed here. 

In general, numerical witness optimization complements rather than replaces analytical criteria. It can identify which occupation probabilities carry the relevant nonclassicality information and thereby guide both their experimental measurement and the construction of new analytical witnesses. This approach is useful because it helps identify where nonclassicality is present and how it varies. Yet, analytical criteria remain particularly valuable in demanding regimes, as illustrated in Fig.~\ref{fig:near_tb}, where a large number of probabilities is required and the numerical optimization consequently becomes increasingly unstable. In this sense, the numerical approach can serve not only as a certification tool, but also as a guide toward simpler and more robust analytical tests.

\vspace{-0.2cm}
\begin{acknowledgments}
 M.\v{Z}. and T.N. acknowledge support from the Czech Science Foundation through Project No. 26-22148S. 
 L.L., D.M. and R.F. acknowledge the project \detokenize{CZ.02.01.01/00/22_008/0004649} (QUEENTEC) of EU and MEYS Czech Republic. 
 P.B. acknowledges support by the research programme of the Strategy AV21 AI: Artificial Intelligence for Science and Society.
 R.F. acknowledges the project No. LUC25006 of MEYS Czech Republic and funding from the European Union's Horizon Europe research and innovation program under grant agreement No 101257526 (SUPERSPIN). This work was supported by the Ministry of Education, Youth and Sports  of the Czech Republic through the e-INFRA CZ (ID:90254).

\end{acknowledgments}


%

\end{document}


\title{Supplemental Material: Nonclassicality from Incomplete Fock-State Statistics}

\author{Martin \v{Z}onda}
\address{Department of Condensed Matter Physics, Faculty of Mathematics and Physics, Charles University, Ke Karlovu 5, Praha 2 CZ-121 16, Czech Republic}

\author{Pavel Bal\'a\v{z}}
\address{FZU -- Institute of Physics of the Czech Academy of Sciences, Na Slovance 1999/2, 182 21 Prague 8, Czech Republic}

\author{Luk\'a\v{s} Lachman}
\address{Department of Optics, Palack\'y University, 17. listopadu 1192/12, 779 00 Olomouc, Czech Republic}

\author{Darren W. Moore}
\address{Department of Optics, Palack\'y University, 17. listopadu 1192/12, 779 00 Olomouc, Czech Republic}

\author{Tom\'a\v{s} Novotn\'y}
\address{Department of Condensed Matter Physics, Faculty of Mathematics and Physics, Charles University, Ke Karlovu 5, Praha 2 CZ-121 16, Czech Republic}

\author{Radim Filip}
\address{Department of Optics, Palack\'y University, 17. listopadu 1192/12, 779 00 Olomouc, Czech Republic}

\date{\today}
\maketitle

\section{Classical probability vectors and linear witnesses~\label{sec:CPW}}

A necessary condition for classicality is that the vector of measured Fock-state probabilities ${P_k}$ can be expressed as a convex combination of Poisson distributions. Consequently, its photon-number distribution is a positive mixture of Poisson distributions,
\begin{equation}
P_k=\int_0^\infty d\mu\,F(\mu)\,p_k(\mu),
\qquad
p_k(\mu)=e^{-\mu}\frac{\mu^k}{k!},
\label{eq:sm_classical_mixture}
\end{equation}
where $F(\mu)\ge0$ is a normalized probability density satisfying
\begin{equation}
\int_0^\infty d\mu\,F(\mu)=1.
\end{equation}

In practice, only a finite subset of occupation probabilities is available. We therefore consider the index set
\begin{equation}
\mathcal{I}=\{k_1,k_2,\ldots,k_M\},
\qquad
k_1<k_2<\cdots<k_M,
\end{equation}
and introduce the vectors
\begin{equation}
\bm{P}_{\mathcal{I}}
=
(P_{k_1},P_{k_2},\ldots,P_{k_M})^{\mathrm T},
\end{equation}
and
\begin{equation}
\bm{p}_{\mathcal{I}}(\mu)
=
(p_{k_1}(\mu),p_{k_2}(\mu),\ldots,p_{k_M}(\mu))^{\mathrm T}.
\end{equation}

Equation~\eqref{eq:sm_classical_mixture} immediately implies
\begin{equation}
\bm{P}_{\mathcal{I}}
=
\int_0^\infty
d\mu\,
F(\mu)
\bm{p}_{\mathcal{I}}(\mu),
\label{eq:sm_convex_representation}
\end{equation}
showing that every classical probability vector belongs to the closed convex hull of the Poisson curve $\mathbf p_I(\mu)$. The closure is important because $\mathbf p_I(\mu)\to\mathbf 0$ as $\mu\to\infty$, while this limiting point need not belong to the convex hull itself. We therefore consider the closed convex set
\begin{equation}
\mathcal C_I=
\overline{\operatorname{conv}}
\left\{\mathbf p_I(\mu):\mu\geq0\right\},
\end{equation}
which contains all classical probability vectors and can be characterized by linear inequalities.

To characterize this convex set we consider an arbitrary linear functional
\begin{equation}
L(\bm{\lambda})
=
\bm{\lambda}^{\mathrm T}
\bm{P}_{\mathcal{I}},
\end{equation}
where
$\bm{\lambda}\in\mathbb{R}^M$.
Using Eq.~\eqref{eq:sm_convex_representation}, one obtains
\begin{align}
L(\bm{\lambda})
&=
\int_0^\infty
d\mu\,
F(\mu)
\,
\bm{\lambda}^{\mathrm T}
\bm{p}_{\mathcal{I}}(\mu)
\nonumber\\
&\le
\int_0^\infty
d\mu\,
F(\mu)
\sup_{\mu'\ge0}
\bm{\lambda}^{\mathrm T}
\bm{p}_{\mathcal{I}}(\mu')
\nonumber\\
&=
\sup_{\mu\ge0}
\bm{\lambda}^{\mathrm T}
\bm{p}_{\mathcal{I}}(\mu).
\label{eq:sm_classical_bound}
\end{align}

Therefore, every classical probability vector satisfies Eq.~\eqref{eq:sm_classical_bound}, with the supremum providing the upper bound of the linear functional over $\mathcal C_I$. This leads to the nonclassicality witness introduced in the main text
\begin{equation}
\mathcal{W}(\bm{\lambda};\bm{P}_{\mathcal{I}})
=
\bm{\lambda}^{\mathrm T}
\bm{P}_{\mathcal{I}}
-
\sup_{\mu\ge0}
\bm{\lambda}^{\mathrm T}
\bm{p}_{\mathcal{I}}(\mu).
\label{eq:sm_witness}
\end{equation}
Every classical state satisfies $\mathcal W\leq0$, so a positive value of the witness certifies nonclassicality.

Conversely, since $\mathcal C_I$ is closed and convex, the separating hyperplane theorem guarantees that every probability vector outside this set can be strictly separated from it by a linear functional. In other words, whenever $\mathbf P_I\notin\mathcal C_I$, there exists a vector $\boldsymbol\lambda$ such that
\begin{equation}
\bm{\lambda}^{\mathrm T}
\bm{P}_{\mathcal{I}}
>
\sup_{\mu\ge0}
\bm{\lambda}^{\mathrm T}
\bm{p}_{\mathcal{I}}(\mu),
\end{equation}
or equivalently,
\begin{equation}
\mathcal{W}(\bm{\lambda};\bm{P}_{\mathcal{I}})>0.
\end{equation}

This establishes the completeness of linear witnesses for the information contained in $\mathbf P_I$: any strict violation of a continuous classicality condition involving only these probabilities places $\mathbf P_I$ outside $\mathcal C_I$ and can therefore be detected by a witness of the form of Eq.~\eqref{eq:sm_witness}. Consequently, optimization over all $\boldsymbol\lambda$ is, in principle, at least as powerful as any of the analytical criteria considered below.

For the numerical implementation, the supplied occupation probabilities must correspond to a physically admissible probability distribution. In particular, all probabilities must be finite and non-negative, and the available probabilities must satisfy
\begin{equation}
\sum_{k\in\mathcal{I}}P_k\le1.
\label{eq:sm_probability_check}
\end{equation}
Violation of this condition implies that the supplied vector cannot originate from any normalized photon-number distribution, irrespective of whether the state is classical or nonclassical.

Finally, the witness is invariant under multiplication of all coefficients by a positive constant,
\begin{equation}
\mathcal{W}(c\bm{\lambda};\bm{P}_{\mathcal{I}})
=
c\,\mathcal{W}(\bm{\lambda};\bm{P}_{\mathcal{I}}),
\qquad c>0,
\label{eq:homo}
\end{equation}
so only the direction of $\bm{\lambda}$ is physically relevant. This freedom is used in the numerical optimization by restricting the coefficients to a bounded interval, which guarantees a compact search domain without changing the set of detectable probability vectors.

The witness formulation presented above admits several analytical solutions that provide simple sufficient conditions for nonclassicality. Throughout this work we use these analytical witnesses both for comparison with the numerical optimization and for constructing the combined analytical criterion ${\cal W}^{\cal K}$ discussed in the main text.

\subsection{Witnesses based on $P_0$ and $P_n$~\label{sec:W0n}}

The most useful analytical family is obtained by considering witnesses involving only the vacuum probability $P_0$ and a single higher occupation probability $P_n$. Optimizing Eq.~\eqref{eq:sm_witness} over both the witness coefficients and the mean occupation number (Poisson parameter) $\mu$ yields a closed classical boundary in the $(P_0,P_n)$ plane.

For $n=1$ all possible choices of the witness coefficients lead to the same condition,
\begin{equation}
P_1>-P_0\ln P_0,
\label{eq:K01}
\end{equation}
which corresponds to the exact classical boundary.

For $n>1$, the optimization produces a piecewise boundary,
\begin{equation}
P_n>
\begin{cases}
\dfrac{P_0}{n!}\left(-\ln P_0\right)^n,
&
0<P_0\le P_0^{\mathrm c},
\\[1.2ex]
C_n(1-P_0),
&
P_0^{\mathrm c}<P_0\le1,
\end{cases}
\label{eq:K0n}
\end{equation}
where the transition point is determined by
\begin{equation}
P_0^{\mathrm c}=e^{-x},
\end{equation}
with
\begin{equation}
x=n+\mathrm{LambertW}\!\left(-ne^{-n}\right),
\label{eq:SMlambert}
\end{equation}
being the non-zero real solution of
\begin{equation}
x=n\left(1-e^{-x}\right),
\end{equation}
and
\begin{equation}
C_n=\frac{x^{n-1}P_0^{\mathrm c}}{(n-1)!}.
\end{equation}

Equation~\eqref{eq:K0n} is equivalent to the construction presented in Ref.~\cite{filip_hierarchy_2013}. Here, however, the transition point is written explicitly in terms of the Lambert $W$ function, which avoids the need for a numerical solution of the transcendental equation. Throughout this work we denote this family of witnesses by ${\cal W}^{0n}$.

\subsection{Klyshko-type criteria}

As a complementary analytical test we employ the well-known Klyshko inequalities,
\begin{equation}
\begin{split}
(k+1)P_{k-1}P_{k+1}-kP_k^2<0,
\qquad k>1,
\\
P_0+\frac{P_1^2}{2P_2}
\left[
\exp\!\left(\frac{2P_2}{P_1}\right)-1
\right]
>1,
\end{split}
\label{eq:SMKlyshko}
\end{equation}
which we denote by ${\cal W}^{{\cal K}_k}$.

Higher-order extensions derived in Ref.~\cite{innocenti_nonclassicality_2022} are used whenever the required probabilities are available,
\begin{equation}
NP_N>
P_{N-1}
\ln\!\left[
\sum_{k=0}^{N-1}
\frac{N^k}{k!}
\frac{P_N^k}{P_{N-1}^k}
+
\frac{N^N}{N!}
\frac{P_N^{N-1}}{P_{N-1}^N}
\left(
1-\sum_{k=0}^{N-2}P_k
\right)
\right],
\label{eq:SMKN}
\end{equation}
which we denote by ${\cal W}^{{\cal K}_N}$.

Whenever applicable, we evaluate all witnesses ${\cal W}^{0n}$, ${\cal W}^{{\cal K}_k}$ and ${\cal W}^{{\cal K}_N}$, and refer to their union as ${\cal W}^{\cal K}$.

\subsection{Majorization hierarchy}

The generalized Klyshko inequalities can be formulated in terms of integer tuples satisfying the majorization relation~\cite{innocenti_nonclassicality_2022}. Given two tuples
\begin{equation}
T^{(0)}\succcurlyeq T^{(1)},
\end{equation}
every classical state satisfies
\begin{equation}
\prod_{i=1}^{s}
T_i^{(1)}!\,
P_{T_i^{(1)}}
\le
\prod_{i=1}^{s}
T_i^{(0)}!\,
P_{T_i^{(0)}}.
\label{eq:SMMajorization}
\end{equation}
Violations define the family of witnesses ${\cal W}^{\cal M}$.

The number of admissible tuples grows rapidly with the number of available probabilities. Rather than constructing the inequalities manually, our implementation generates them automatically. Starting from the set of experimentally available photon numbers, all integer tuples with equal length and equal total photon number are enumerated. Every pair satisfying the majorization relation is retained, after which two additional filters are applied.

First, tuple pairs sharing common photon numbers are removed, since they either lead to trivial identities or to inequalities equivalent to lower-order ones. Second, pairs whose multiplicities possess a common integer divisor are discarded, eliminating inequalities that are simple integer multiples of already existing ones. The remaining primitive tuple pairs are then converted into inequalities of the form of Eq.~\eqref{eq:SMMajorization} and evaluated independently. The overall majorization criterion is satisfied whenever at least one inequality is violated. For consecutive photon numbers up to $N_P=8$, these primitive tuple pairs are precomputed and stored, while for arbitrary subsets of measured probabilities they are generated on demand. 

For example, choosing
\[
T^{(0)}=(k,k),
\qquad
T^{(1)}=(k-1,k+1),
\]
immediately reproduces the standard Klyshko inequality, while
\[
T^{(0)}=(k+1,k+2),
\qquad
T^{(1)}=(k,k+3),
\]
yields the four-probability condition
\begin{equation}
P_{k+1}P_{k+2}
\le
\frac{k+3}{k+1}
P_kP_{k+3},
\end{equation}
illustrating how the hierarchy naturally generates increasingly non-local inequalities.

\subsection{Additional two-probability criteria~\label{sec:Wadd}}

Reference~\cite{innocenti_nonclassicality_2022} also derived implicit boundaries involving only two arbitrary non-zero photon-number probabilities,
\begin{equation}
\left(k!P_k\right)^{\frac{m}{m-k}}
=
\exp\!\left[
-\left(
\frac{m!P_m}{k!P_k}
\right)^{\frac{1}{m-k}}
\right]
\left(m!P_m\right)^{\frac{k}{m-k}}.
\label{eq:SM2Pk}
\end{equation}

However, Eq.~\eqref{eq:SM2Pk} does not generally describe the complete classical boundary in the $(P_k,P_m)$ plane. It corresponds only to the Poisson branch itself, whereas the physically relevant classical boundary is given by the convex hull of this branch. A simple example is provided by $k=0$ and $m>1$, where Eq.~\eqref{eq:SM2Pk} reduces to
\[
P_m=\frac{P_0}{m!}
\left(-\ln P_0\right)^m,
\]
which reproduces only the first branch of Eq.~\eqref{eq:K0n}. The complete classical boundary additionally contains the linear segment obtained by convexification.

For this reason, it is generally not possible to determine a priori whether Eq.~\eqref{eq:SM2Pk} represents the full classical boundary for a given pair $(k,m)$. Throughout this work these two-probability criteria are therefore \emph{not} included among the analytical witnesses used for comparison with the numerical optimization. They are discussed separately later in Sec.~\ref{sec:2prob} of this Supplemental Material (SM), where their range of applicability is illustrated.

\subsection{Counting-matrix criteria}

The final family of analytical criteria considered in this work is based on counting matrices constructed directly from the occupation probabilities~\cite{krishnaswamy2026observable}. We briefly summarize the construction here and explain how the admissible matrices are generated in
our implementation.

For a classical state, the normally ordered expectation value
\begin{equation}
\left\langle :\hat f^\dagger \hat f: \right\rangle \geq 0
\end{equation}
must be non-negative for an arbitrary operator $\hat f$. As shown in
Ref.~\cite{krishnaswamy2026observable}, choosing an appropriate expansion of
$\hat f$ and using the photoelectric counting relation
\begin{equation}
P_m=
\left\langle :
\frac{\hat n^m}{m!}e^{-\hat n}
:\right\rangle
\end{equation}
leads to the counting matrix
\begin{equation}
C^{(\mathcal I)}_{kl}=(k+l)!\,P_{k+l},
\qquad k,l\in\mathcal I .
\label{eq:SM_counting_matrix}
\end{equation}
For every classical state this matrix must be positive semidefinite,
\begin{equation}
C^{(\mathcal I)}\succeq0.
\label{eq:SM_count_psd}
\end{equation}
Consequently, a negative eigenvalue of any admissible counting matrix
certifies nonclassicality. We denote the resulting family of criteria by
${\cal W}^{\cal C}$.

The index set $\mathcal I$ must be chosen such that $k+l$ is a non-negative
integer corresponding to an available probability for every pair
$k,l\in\mathcal I$. The indices within a given matrix can therefore be either
all integers or all half-integers~\cite{krishnaswamy2026observable}. For
example,
\begin{equation}
\mathcal I=\{0,1\}
\end{equation}
gives
\begin{equation}
C^{(0,1)}
=
\begin{pmatrix}
P_0 & P_1\\
P_1 & 2P_2
\end{pmatrix}.
\label{eq:SM_C01}
\end{equation}
Positive semidefiniteness requires
\begin{equation}
2P_0P_2-P_1^2\geq0,
\end{equation}
and its violation reproduces the corresponding Klyshko criterion.

The half-integer construction generates the complementary matrices. For
example,
\begin{equation}
\mathcal I=
\left\{\frac{1}{2},\frac{3}{2}\right\}
\end{equation}
gives
\begin{equation}
C^{(1/2,3/2)}
=
\begin{pmatrix}
P_1 & 2P_2\\
2P_2 & 6P_3
\end{pmatrix},
\label{eq:SM_Chalf}
\end{equation}
and hence
\begin{equation}
6P_1P_3-4P_2^2\geq0.
\end{equation}
Its violation,
\begin{equation}
2P_2^2>3P_1P_3,
\end{equation}
again reproduces the corresponding Klyshko criterion. More generally, a
two-element index set $\mathcal I=\{k,l\}$ gives
\begin{equation}
C^{(k,l)}
=
\begin{pmatrix}
(2k)!\,P_{2k} & (k+l)!\,P_{k+l}\\
(k+l)!\,P_{k+l} & (2l)!\,P_{2l}
\end{pmatrix}.
\label{eq:SM_Ckl}
\end{equation}
Thus, a nontrivial $2\times2$ counting matrix requires three occupation
probabilities whose indices $2k$, $k+l$, and $2l$ form an arithmetic
progression.

Larger matrices combine information from more probabilities. For example,
if $P_0,\ldots,P_4$ are available, the integer index set
$\mathcal I=\{0,1,2\}$ gives
\begin{equation}
C^{(0,1,2)}
=
\begin{pmatrix}
P_0  & P_1  & 2P_2\\
P_1  & 2P_2 & 6P_3\\
2P_2 & 6P_3 & 24P_4
\end{pmatrix},
\label{eq:SM_C012}
\end{equation}
while the half-integer set
$\mathcal I=\{1/2,3/2\}$ gives Eq.~\eqref{eq:SM_Chalf}. Both matrices must be
positive semidefinite for a classical state.

For a general set of available occupation indices $\mathcal P$, we call an
index set $\mathcal I$ admissible if
\begin{equation}
k+l\in\mathcal P
\qquad
\text{for all } k,l\in\mathcal I.
\label{eq:SM_count_admissible}
\end{equation}
Our implementation constructs the admissible integer and half-integer index
sets and retains the maximal ones. It is sufficient to test only these
matrices: a matrix associated with a subset of a maximal index set is a
principal submatrix of the corresponding maximal matrix. If the maximal
matrix is positive semidefinite, all its principal submatrices are positive
semidefinite as well. Their automatic construction and evaluation are included in the accompanying open-source
code~\cite{code_github}.

\subsection{Nonclassical Characteristic Functions}

As an independent benchmark, we also evaluate nonclassicality using the
quadrature characteristic function. We define the phase-dependent quadrature and its characteristic
function as
\begin{equation}
    \hat{x}_{\theta}
    =
    \frac{1}{\sqrt{2}}
    \left(
    \hat{a}e^{-i\theta}
    +
    \hat{a}^{\dagger}e^{i\theta}
    \right),
    \qquad
    \phi_{\theta}(t)
    =
    \mathrm{Tr}
    \left[
    \rho\,e^{it\hat{x}_{\theta}}
    \right].
\end{equation}
For any classical state, the modulus of the quadrature characteristic
function is bounded by that of the vacuum state~\cite{Vogel2000nonclassical},
\begin{equation}
    |\phi_{\theta}(t)|
    \leq
    |\phi_0(t)|
    =
    e^{-t^2/4}.
\end{equation}
Consequently, the existence of a quadrature angle $\theta$ and
$t\in\mathbb{R}$ such that
\begin{equation}
    |\phi_{\theta}(t)|>e^{-t^2/4}
    \label{eq:cf_nonclass}
\end{equation}
is sufficient to certify nonclassicality. This corresponds to the
first-order characteristic-function criterion; a complete hierarchy of
necessary and sufficient conditions was developed in
Ref.~\cite{Richter2002nonclassicality}.

For the Jaynes--Cummings dynamics considered in the main text, the oscillator is initially
thermal and its reduced density matrix remains diagonal in the Fock basis,
\begin{equation}
    \rho(\tau)
    =
    \sum_{n=0}^{\infty}
    P_n(\tau)\ket{n}\bra{n}.
\end{equation}

Unlike the criteria based on a finite subset of occupation probabilities, the characteristic-function test requires the full photon-number statistics, since all $P_n(\tau)$ contribute to the sum below. In numerical calculations this means that the Fock-space cutoff must be chosen sufficiently large that the omitted tail is negligible.

Such a state is invariant under phase rotations, and therefore
$\phi_{\theta}(t)$ is independent of the quadrature angle $\theta$. The
characteristic function can then be written as a statistical mixture of the
Fock-state characteristic functions,
\begin{equation}
    \phi(t)
    =
    \sum_{n=0}^{\infty}
    P_n(\tau)\phi_n(t),
\end{equation}
where
\begin{equation}
    \phi_n(t)
    =
    e^{-t^2/4}
    \mathcal{L}_n\left(\frac{t^2}{2}\right)
    =
    \phi_0(t)
    \mathcal{L}_n\left(\frac{t^2}{2}\right),
\end{equation}
and $\mathcal{L}_n(x)$ denotes the Laguerre polynomial of order $n$. Equation~\eqref{eq:cf_nonclass}
therefore reduces to
\begin{equation}
    \left|
    \sum_{n=0}^{\infty}
    P_n(\tau)
    \mathcal{L}_n\left(\frac{t^2}{2}\right)
    \right|>1
\end{equation}
for at least one $t\in\mathbb{R}$.

\subsection{Semi-infinite linear-program formulation}
\label{sec:SM_silp}

We now describe the numerical procedure used to decide whether a finite vector of occupation probabilities is compatible with a classical state. The starting point is the witness condition
\begin{equation}
{\cal W}(\bm{\lambda};\bm{P}_{\mathcal I})
=
\bm{\lambda}^{\mathrm T}\bm{P}_{\mathcal I}
-
\sup_{\mu\ge0}
\bm{\lambda}^{\mathrm T}\bm{p}_{\mathcal I}(\mu),
\label{eq:SM_witness_num}
\end{equation}
where
\begin{equation}
\bm{P}_{\mathcal I}
=
(P_{k_1},\ldots,P_{k_M})^{\mathrm T},
\qquad
\bm{p}_{\mathcal I}(\mu)
=
(p_{k_1}(\mu),\ldots,p_{k_M}(\mu))^{\mathrm T}.
\end{equation}
The probability vector is certified as nonclassical if there exists a real vector
$\bm{\lambda}$ such that
\begin{equation}
{\cal W}(\bm{\lambda};\bm{P}_{\mathcal I})>0.
\end{equation}

A direct numerical approach would be to maximize
${\cal W}$ with respect to $\bm{\lambda}$, or equivalently to minimize
\begin{equation}
f(\bm{\lambda})
=
-{\cal W}(\bm{\lambda};\bm{P}_{\mathcal I})
=
\sup_{\mu\ge0}
\bm{\lambda}^{\mathrm T}\bm{p}_{\mathcal I}(\mu)
-
\bm{\lambda}^{\mathrm T}\bm{P}_{\mathcal I}.
\label{eq:SM_direct_objective}
\end{equation}
This form is inconvenient numerically. The function is convex in
$\bm{\lambda}$, but it is generally nonsmooth because it contains a supremum over $\mu$. As $\bm{\lambda}$ changes, the supremum may be attained at different local maxima or approached only in the limit $\mu\rightarrow\infty$. In practice, this makes gradient-based minimization unstable and derivative-free methods unnecessarily sensitive to the initial point.

We therefore rewrite the problem in a form that keeps its exact structure but avoids direct optimization of the non-smooth supremum. We introduce an auxiliary variable $t$ constrained by
\begin{equation}
t\ge
\bm{\lambda}^{\mathrm T}\bm{p}_{\mathcal I}(\mu)
\qquad
\text{for all } \mu\ge0.
\label{eq:SM_t_constraint}
\end{equation}
For fixed $\bm{\lambda}$, the smallest admissible value of $t$ is therefore given by
\begin{equation}
t=
\sup_{\mu\ge0}
\bm{\lambda}^{\mathrm T}\bm{p}_{\mathcal I}(\mu),
\end{equation}
where the supremum includes the limiting value at $\mu\rightarrow\infty$. Hence minimizing
Eq.~\eqref{eq:SM_direct_objective} is equivalent to the optimization problem
\begin{align}
\min_{\bm{\lambda},t}\quad
&
t-\bm{\lambda}^{\mathrm T}\bm{P}_{\mathcal I},
\label{eq:SM_silp_objective}
\\
\text{subject to}\quad
&
t\ge
\bm{\lambda}^{\mathrm T}\bm{p}_{\mathcal I}(\mu),
\qquad
\forall\,\mu\ge0,
\label{eq:SM_silp_constraint_mu}
\\
&
t\ge0,
\label{eq:SM_silp_constraint_t}
\\
&
-1\le\lambda_j\le1,
\qquad
j=1,\ldots,M.
\label{eq:SM_silp_constraint_lambda}
\end{align}
The constraint $t\ge0$ is important. Since
$\bm{p}_{\mathcal I}(\mu)\to0$ for $\mu\to\infty$, the supremum of
$\bm{\lambda}^{\mathrm T}\bm{p}_{\mathcal I}(\mu)$ is always at least zero, even if it is not attained at a finite value of $\mu$. The condition $t\ge0$ therefore accounts for the point at infinity.

The bounds on $\lambda_j$ do not restrict the set of detectable states. The witness is homogeneous, Eq.~\eqref{eq:homo}, so only the direction of $\bm{\lambda}$ matters. The bounds merely fix a convenient normalization and make the feasible domain compact.

For any fixed $\mu$, the constraint
Eq.~\eqref{eq:SM_silp_constraint_mu} is linear in the unknowns
$(\bm{\lambda},t)$, and the objective function
Eq.~\eqref{eq:SM_silp_objective} is linear as well. The only complication is that the number of constraints is infinite, one for each $\mu\ge0$. The problem is therefore a semi-infinite linear program.

We solve it by an exchange, or cutting-plane, procedure. One starts from a finite set of values
\begin{equation}
{\cal M}=\{\mu_1,\ldots,\mu_R\},
\end{equation}
and solves the finite linear program
\begin{align}
\min_{\bm{\lambda},t}\quad
&
t-\bm{\lambda}^{\mathrm T}\bm{P}_{\mathcal I},
\label{eq:SM_finite_lp_objective}
\\
\text{subject to}\quad
&
t\ge
\bm{\lambda}^{\mathrm T}\bm{p}_{\mathcal I}(\mu_r),
\qquad
\mu_r\in{\cal M},
\label{eq:SM_finite_lp_constraint_mu}
\\
&
t\ge0,
\\
&
-1\le\lambda_j\le1.
\end{align}
This gives a candidate solution $(\bm{\lambda},t)$ satisfying all constraints in the current finite set.

The candidate is then checked against the full continuum of constraints. For the obtained vector $\bm{\lambda}$ we compute
\begin{equation}
w=
\sup_{\mu\ge0}
g(\mu),
\qquad
g(\mu)=
\bm{\lambda}^{\mathrm T}\bm{p}_{\mathcal I}(\mu).
\label{eq:SM_worst_case}
\end{equation}
If
\begin{equation}
w\le t+\delta,
\label{eq:SM_cut_tolerance}
\end{equation}
where $\delta$ is the numerical tolerance, the current solution is accepted. Otherwise the maximizing value $\mu^\ast$ defines the most violated constraint,
\begin{equation}
t\ge
\bm{\lambda}^{\mathrm T}\bm{p}_{\mathcal I}(\mu^\ast),
\end{equation}
and this point is added to the working set,
\begin{equation}
{\cal M}\cup\{\mu^\ast\}\rightarrow{\cal M}.
\end{equation}
The finite linear program is then solved again. The procedure is repeated until Eq.~\eqref{eq:SM_cut_tolerance} is fulfilled.

It remains to specify how the one-dimensional maximization in
Eq.~\eqref{eq:SM_worst_case} is carried out. It is useful to extend the coefficients $\lambda_j$ to a full sequence $\Lambda_k$ by setting
\begin{equation}
\Lambda_k=
\begin{cases}
\lambda_j, & k=k_j\in{\cal I},\\
0, & k\notin{\cal I}.
\end{cases}
\end{equation}
Then
\begin{equation}
g(\mu)
=
e^{-\mu}
\sum_{k=0}^{K}
\Lambda_k\frac{\mu^k}{k!},
\label{eq:SM_gmu}
\end{equation}
where $K=\max{\cal I}$. Its derivative is
\begin{align}
\frac{d g}{d\mu}
&=
e^{-\mu}
\left[
\sum_{k=1}^{K}
\Lambda_k\frac{\mu^{k-1}}{(k-1)!}
-
\sum_{k=0}^{K}
\Lambda_k\frac{\mu^k}{k!}
\right]
\nonumber\\
&=
e^{-\mu}
\sum_{k=0}^{K}
(\Lambda_{k+1}-\Lambda_k)
\frac{\mu^k}{k!},
\label{eq:SM_gmu_derivative}
\end{align}
with $\Lambda_{K+1}=0$. Thus the finite stationary points are obtained from the polynomial equation
\begin{equation}
\sum_{k=0}^{K}
(\Lambda_{k+1}-\Lambda_k)
\frac{\mu^k}{k!}
=0.
\label{eq:SM_stationary_polynomial}
\end{equation}
In the implementation we find all roots of this polynomial, retain only the real non-negative ones, and add the boundary point $\mu=0$. The value of $g(\mu)$ is evaluated on this finite candidate set. The limiting value
\begin{equation}
\lim_{\mu\to\infty}g(\mu)=0
\end{equation}
is included through the constraint $t\ge0$. Including this limiting value, the supremum over $\mu\geq 0$ can be evaluated exactly as the maximum over a finite set of candidate values
\begin{equation}
w=
\max\left[
0,\,
g(0),\,
g(\mu_1^\ast),\,
g(\mu_2^\ast),\ldots
\right],
\label{eq:SM_worst_case_eval}
\end{equation}
where $\{\mu_i^\ast\}$ are the non-negative real roots of
Eq.~\eqref{eq:SM_stationary_polynomial}.

After convergence of the exchange procedure, the optimized value
\begin{equation}
f_{\min}
=
t-\bm{\lambda}^{\mathrm T}\bm{P}_{\mathcal I}
\end{equation}
is the minimum of Eq.~\eqref{eq:SM_direct_objective}. Equivalently, the optimized witness value is
\begin{equation}
{\cal W}_{\max}
=
-f_{\min}
=
\bm{\lambda}^{\mathrm T}\bm{P}_{\mathcal I}-t.
\end{equation}
The probability vector is classified as nonclassical if
\begin{equation}
{\cal W}_{\max}>\varepsilon,
\end{equation}
where $\varepsilon$ is the numerical tolerance used to avoid decisions based on round-off errors.

In summary, the numerical criterion consists of the following steps. First, a finite linear program is solved for a current set of Poisson parameters ${\cal M}$. Second, the resulting witness is checked by evaluating the supremum over $\mu\geq0$, including the limiting value at infinity. If the supremum exceeds the current bound, it is attained at a finite value of $\mu$, which is then added to $\mathcal M$. This preserves the original witness condition while replacing the direct non-smooth optimization by a sequence of stable linear programs.

\subsubsection{Illustration of the procedure}
\label{sec:SM_exchange_example}

To illustrate the role of the auxiliary variable $t$ and the exchange
procedure, we consider the simplest case in which only $P_0$ and $P_1$
are used. We choose the probability vector
\begin{equation}
\bm P=(P_0,P_1)=(0.5,0.4).
\end{equation}
For comparison, the exact classical boundary in this plane is
\begin{equation}
P_1=-P_0\ln P_0\simeq0.3466,
\end{equation}
so this point is nonclassical.

The numerical method does not use this analytical boundary. It starts from
the general optimization problem
\begin{equation}
\min_{\lambda_0,\lambda_1,t}
\left(t-\lambda_0P_0-\lambda_1P_1\right),
\end{equation}
subject to
\begin{equation}
t\geq
e^{-\mu}\left(\lambda_0+\lambda_1\mu\right),
\qquad \forall\,\mu\geq0,
\end{equation}
together with $t\geq0$ and
$-1\leq\lambda_0,\lambda_1\leq1$.

We begin with the minimal working set
\begin{equation}
\mathcal M_0=\{0\}.
\end{equation}
The only Poisson constraint is then
\begin{equation}
t\geq\lambda_0.
\end{equation}
For the chosen positive probabilities, the finite linear program is minimized
by
\begin{equation}
\lambda_0=0,\qquad
\lambda_1=1,\qquad
t=0.
\end{equation}
This is only a candidate solution because the constraint has so far been
checked at $\mu=0$ only.

For this candidate,
\begin{equation}
g(\mu)
=
\bm\lambda\cdot\bm p(\mu)
=
\mu e^{-\mu}.
\end{equation}
Its maximum occurs at
\begin{equation}
\mu^\ast=1,
\qquad
w=g(1)=e^{-1}.
\end{equation}
Since $w>t$, the candidate violates the omitted constraint at $\mu=1$.
We therefore enlarge the working set to
\begin{equation}
\mathcal M_1=\{0,1\}.
\end{equation}

The second finite linear program contains the constraints
\begin{equation}
t\geq\lambda_0,
\qquad
t\geq e^{-1}(\lambda_0+\lambda_1).
\end{equation}
Its solution is
\begin{equation}
\lambda_1=1,
\qquad
\lambda_0=\frac{1}{e-1}\simeq0.582,
\qquad
t=\frac{1}{e-1}.
\end{equation}
The corresponding continuous function is now
\begin{equation}
g(\mu)
=
e^{-\mu}
\left(
\frac{1}{e-1}+\mu
\right).
\end{equation}
Its maximum occurs at
\begin{equation}
\mu^\ast
=
1-\frac{1}{e-1}
\simeq0.418,
\end{equation}
where
\begin{equation}
w\simeq0.658>t\simeq0.582.
\end{equation}
Thus the constraint at $\mu^\ast\simeq0.418$ is added and the linear
program is solved again.

Repeating this procedure successively adds the values of $\mu$ at which the
current candidate violates the classical bound most strongly. For the present
example the first iterations are
\begin{equation}
\begin{array}{c|cccc}
\text{iteration}
&
\lambda_0
&
t
&
\mu^\ast
&
w-t
\\ \hline
0 & 0       & 0       & 1.000 & 3.6788\times10^{-1}\\
1 & 0.58198 & 0.58198 & 0.41802 & 7.6370\times10^{-2}\\
2 & 0.31906 & 0.48525 & 0.68094 & 2.0885\times10^{-2}\\
3 & 0.16800 & 0.42968 & 0.83200 & 5.4946\times10^{-3}\\
4 & 0.24543 & 0.46887 & 0.75457 & 1.3389\times10^{-3}\\
5 & 0.28269 & 0.48773 & 0.71731 & 3.3060\times10^{-4}\\
6 & 0.30098 & 0.49699 & 0.69902 & 8.2148\times10^{-5}\\
\vdots & \vdots & \vdots & \vdots & \vdots\\
\infty & 0.30685 & 0.50000 & 0.69315 & 0
\end{array}
\label{eq:SM_exchange_iterations}
\end{equation}

The individual candidate parameters need not approach their limiting values
monotonically, since each newly added constraint changes the active set of
the finite linear program. Nevertheless, the maximal constraint violation
$w-t$ decreases rapidly in this example. The limiting values are
$\lambda_0^\ast=1-\ln 2$, $t^\ast=1/2$, and
$\mu^\ast=\ln 2$ and agree with the exact solution that can be obtained analytically: For
$\lambda_1=1$,
\begin{equation}
g(\mu)
=
e^{-\mu}(\lambda_0+\mu)
\end{equation}
has its maximum at
\begin{equation}
\mu^\ast=1-\lambda_0,
\end{equation}
provided $\lambda_0<1$. Hence
\begin{equation}
t=e^{\lambda_0-1}.
\end{equation}
The remaining optimization is therefore
\begin{equation}
\min_{\lambda_0}
\left[
e^{\lambda_0-1}
-\lambda_0P_0-P_1
\right].
\end{equation}
The stationary condition gives
\begin{equation}
e^{\lambda_0-1}=P_0,
\end{equation}
or
\begin{equation}
\lambda_0^\ast=1+\ln P_0.
\end{equation}
At the optimum,
\begin{equation}
t^\ast=P_0,
\end{equation}
and the optimized witness is
\begin{align}
{\cal W}_{\max}
&=
\lambda_0^\ast P_0+P_1-t^\ast
\nonumber\\
&=
P_1+P_0\ln P_0.
\end{align}
Therefore,
\begin{equation}
{\cal W}_{\max}>0
\end{equation}
is equivalent to
\begin{equation}
P_1>-P_0\ln P_0,
\end{equation}
which is exactly the known classical boundary.

\section{Illustrative examples}

In this section we illustrate the numerical optimization procedure using probability vectors without reference to a specific physical model. These examples serve two purposes. First, they visualize the geometry of the classical set introduced in Sec.~\ref{sec:SM_silp}. Second, they provide a direct comparison between the numerical optimization and the analytical criteria discussed above. 

\subsection{Complete set of two-probability criteria \label{sec:2prob}}
\begin{figure}[ht!]
	\includegraphics[width=1.0\columnwidth]{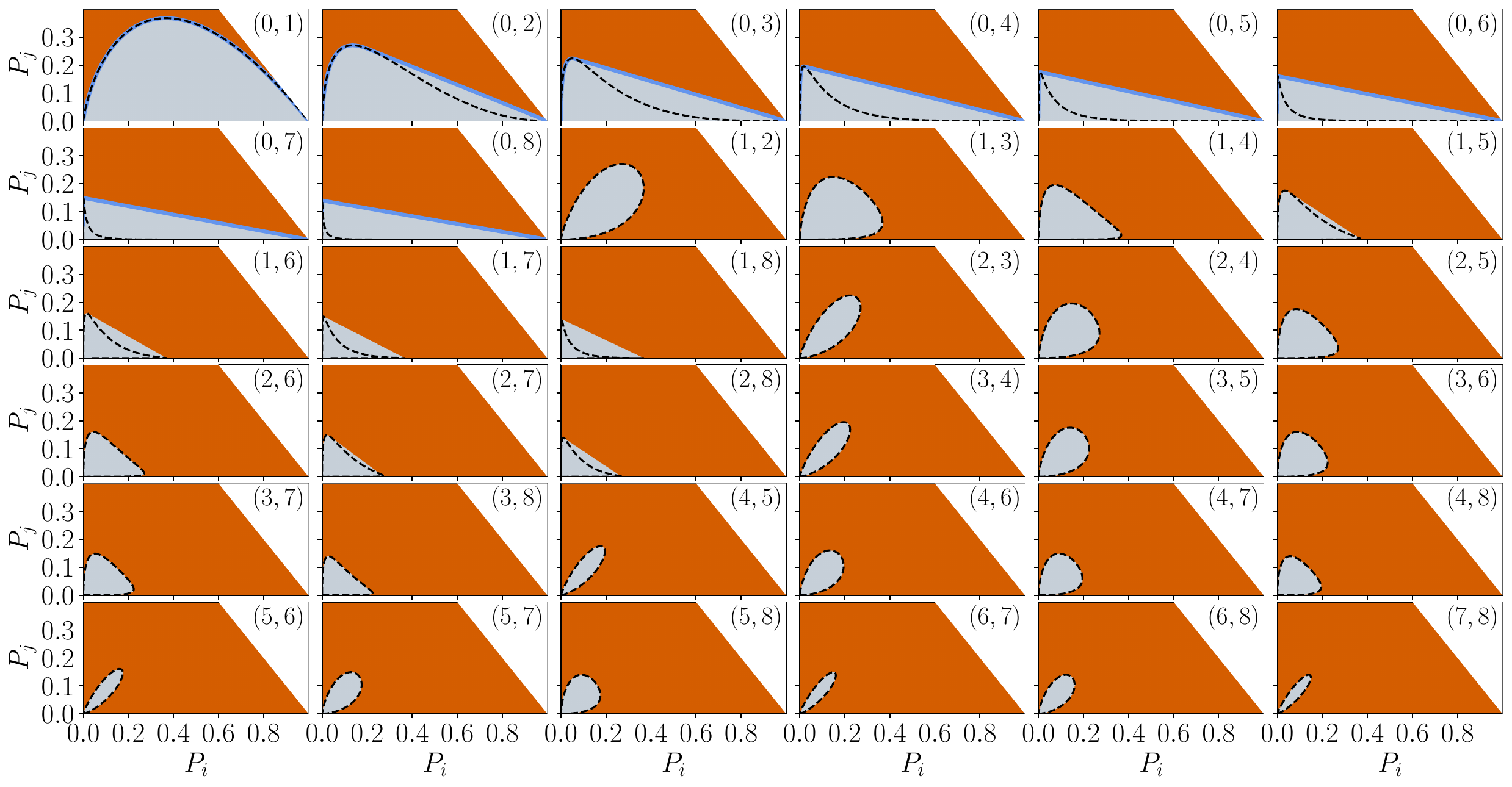}\caption{
Comparison of nonclassical regions obtained from the numerical criterion with classicality boundaries in the reduced space of two probabilities $(P_i,P_j)$, where each subplot corresponds to a specific pair $(i,j)$ with $j\leq 8$. The orange regions indicate nonclassical states identified by the numerical optimization. The blue solid curves show the corresponding analytical boundaries given by Eqs.~\eqref{eq:K01} and~\eqref{eq:K0n} (${\cal W}^{0n}$). The dashed black curves represent the two-probability boundaries given by Eq.~\eqref{eq:SM2Pk}. As discussed in the text, the latter curves coincide with the true classical boundary only in cases where they define a convex envelope; otherwise, they capture only a subset of the classical boundary. White region shows where condition $P_i+P_j\leq 1$ is violated.
\label{fig:all2}}
\end{figure}

As a first validation of the numerical procedure described in Sec.~\ref{sec:SM_silp}, we investigate nonclassicality criteria involving only two occupation probabilities. Figure~\ref{fig:all2} presents the numerical nonclassicality maps for all combinations $(P_i,P_j)$ with $i,j\leq8$. These maps are model-independent and directly characterize the geometry of the classical set. Consequently, they can be applied to any physical system for which the corresponding occupation probabilities are available.

The orange regions indicate probability vectors identified as nonclassical by the numerical optimization. For all cases involving the vacuum probability, $(P_0,P_j)$, the blue solid curves show the analytical boundaries of Eqs.~\eqref{eq:K01}~\eqref{eq:K0n}, corresponding to the witnesses ${\cal W}^{0n}$ from Sec.~\ref{sec:W0n}. The numerical boundaries are indistinguishable from the analytical ones within numerical precision. These panels extend the example shown in the Figure~1 of the main text to all two-probability witnesses involving $P_j$ up to $j=8$.

For comparison, the dashed black curves denote the implicit two-probability boundaries of Eq.~\eqref{eq:SM2Pk}. Unlike Eq.~\eqref{eq:K0n}, these expressions are available for arbitrary pairs $(P_i,P_j)$ and are therefore shown in every panel. However, as already discussed~\cite{innocenti_nonclassicality_2022}, they represent the true classical boundary only when they already coincide with the convex envelope of the corresponding Poisson curve. Otherwise, they reproduce only the nonlinear branch and fail to describe the linear segment generated by convexification. The comparison with the numerical results clearly illustrates this limitation. Whenever Eq.~\eqref{eq:SM2Pk} does not define the convex envelope, the dashed curve lies entirely inside the classical region determined by the numerical optimization. Consequently, Eq.~\eqref{eq:SM2Pk} generally overestimates the nonclassical region and is therefore not included among the analytical criteria used throughout the main text and in the code.

\begin{figure}[!ht]
    \includegraphics[width=1.0\columnwidth]{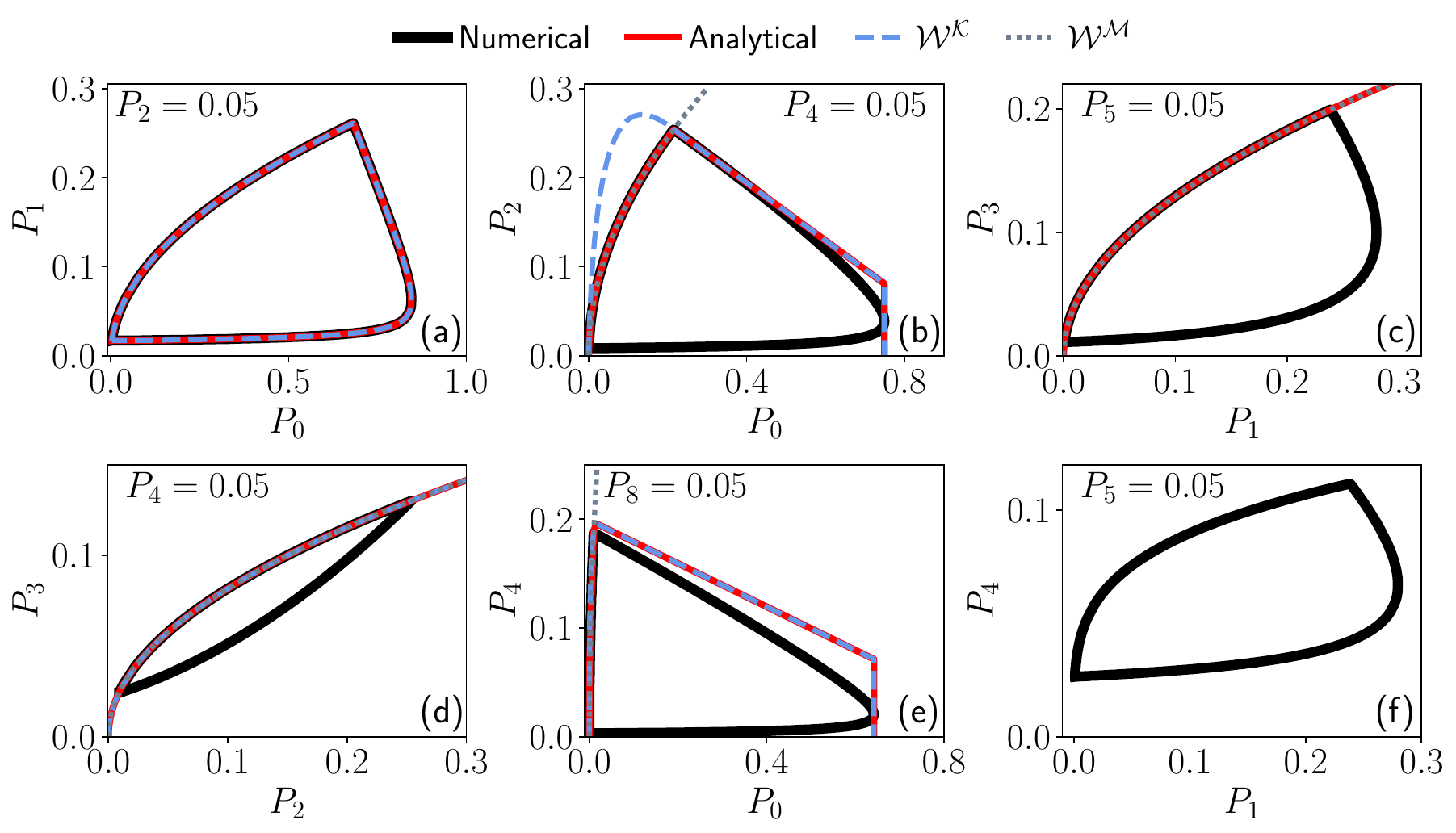}
    \caption{
    Comparison of nonclassicality boundaries obtained from analytical (red) and numerical (black) criteria for selected examples in which only three occupation probabilities are known with sufficient precision. Blue dashed curves correspond to boundaries derived from the witnesses ${\cal W}^{\cal K}$, while dotted gray curves indicate boundaries obtained from the majorization witnesses ${\cal W}^{\cal M}$.
    }
\label{fig:3Pex}
\end{figure}
\subsection{Tests of criteria based on three and four probabilities}
Figure~\ref{fig:3Pex} presents several representative examples of nonclassicality boundaries obtained from the numerical optimization (black curves) and from the available analytical criteria. The analytical results are constructed from all applicable witnesses ${\cal W}^{0n}$ [Eq.~\eqref{eq:K0n}], ${\cal W}^{{\cal K}_k}$ [Eq.~\eqref{eq:SMKlyshko}], and majorization witnesses ${\cal W}^{\cal M}$. For clarity, the boundaries generated by the Klyshko-type witnesses are shown as blue dashed curves, while those obtained from majorization criteria are shown as gray dotted curves. The resulting analytical boundary, obtained by combining all available witnesses, is displayed as a solid red curve.

The selected examples illustrate several important features. Panel (a) shows the case where three consecutive probabilities, $P_0$, $P_1$, and $P_2$, are available. In this situation, the standard Klyshko witness is sufficient to reproduce exactly the same boundary as obtained from the numerical optimization. We have not found any counterexample within the $(P_0,P_1,P_2)$ probability space for which the numerical boundary extends beyond that predicted by the Klyshko criterion. However, this property does not hold for arbitrary triples of consecutive probabilities. An example for $(P_2,P_3,P_4)$ is shown in panel (d), where the analytical criteria recover only the upper branch of the lens-shaped boundary obtained numerically. Consequently, the region identified as nonclassical by the numerical approach is substantially larger than that detected by the available analytical witnesses.

Panels (b), (c), and (e) demonstrate situations involving higher-order occupation probabilities and illustrate the usefulness of the majorization witnesses ${\cal W}^{\cal M}$. In these cases, the majorization criteria allow the analytical boundaries to follow the numerical ones over significant portions of the parameter space. Panel (c) is particularly noteworthy because no Klyshko witness is available: the probability $P_0$ is not known and the measured probabilities are not consecutive. Nevertheless, the majorization criteria still detect nonclassicality in part of the accessible probability space.

\begin{figure}[!ht]
    \includegraphics[width=1.0\columnwidth]{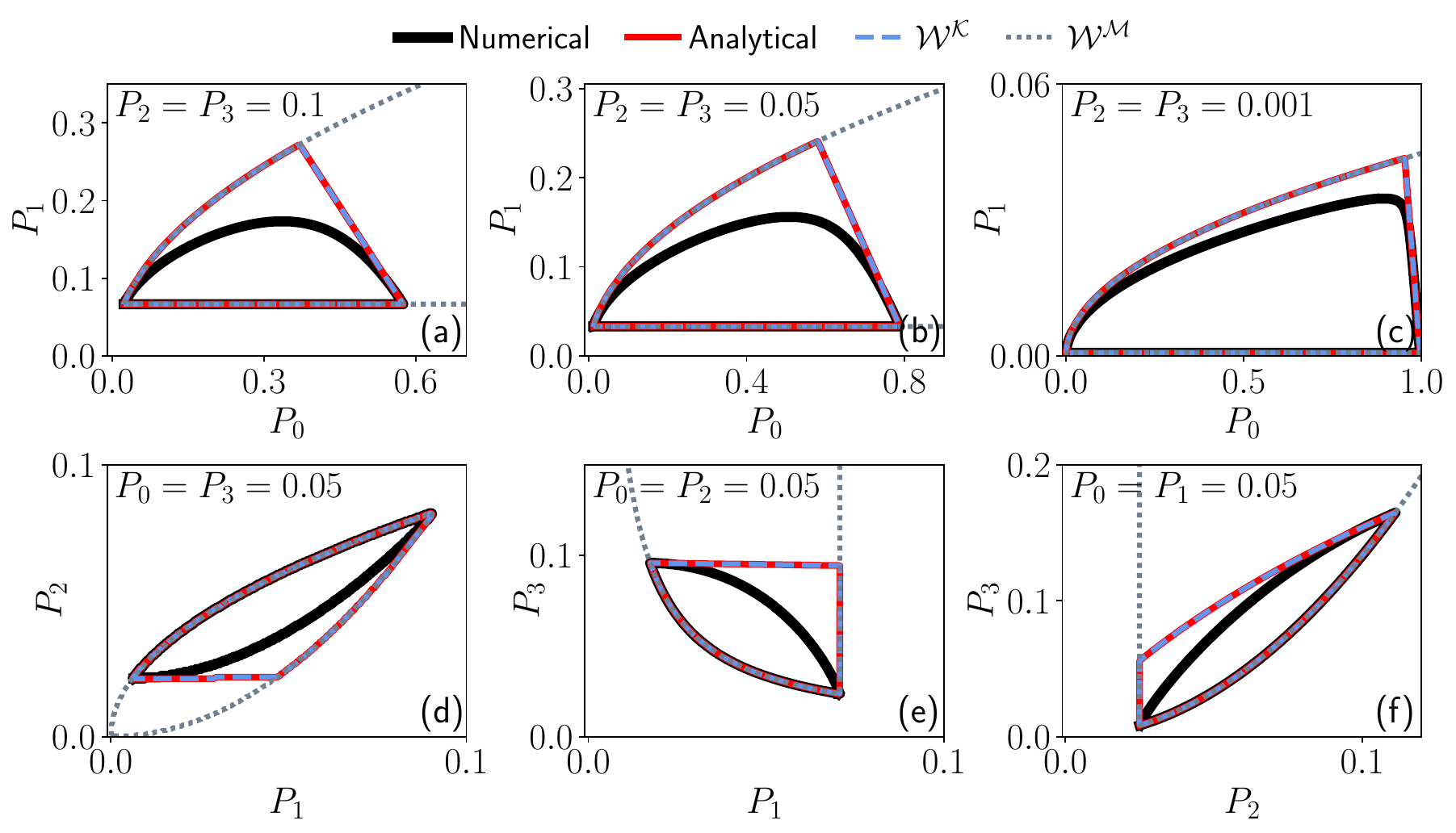}
    \caption{
    Comparison of nonclassicality boundaries obtained from analytical (red) and numerical (black) criteria for selected examples in which four occupation probabilities are known with sufficient precision. Blue dashed curves correspond to boundaries derived from the witnesses ${\cal W}^{\cal K}$, while dotted gray curves indicate boundaries obtained from the majorization witnesses ${\cal W}^{\cal M}$.
    }
\label{fig:4Pex}
\end{figure}
Finally, panel (f) highlights the main advantage of the numerical approach. In this example, neither Klyshko nor majorization witnesses can be constructed from the available probabilities. Despite the absence of any analytical criterion, the numerical optimization reveals that a large fraction of the $(P_1,P_4)$ plane for $P_7=0.05$ corresponds to nonclassical states. This example demonstrates that the numerical criteria can remain highly effective even when only a small number of occupation probabilities are known with high precision.

In the following sections, where we investigate specific physical models, we focus primarily on situations in which all probabilities up to some excitation number $N$ are available. Figure~\ref{fig:4Pex} indicates that, when four occupation probabilities are known, the numerical criteria typically cover a substantially larger fraction of the probability space than the analytical ones. In addition, we have observed, that if we use all probabilities up to $N$, there is no advantage in using ${\cal W}^{\cal M}$ if ${\cal W}^{\cal K}$ have already been used.  

\pagebreak

\subsection{Criteria based on counting matrix}
\begin{figure}[!ht]
    \includegraphics[width=1.0\columnwidth]{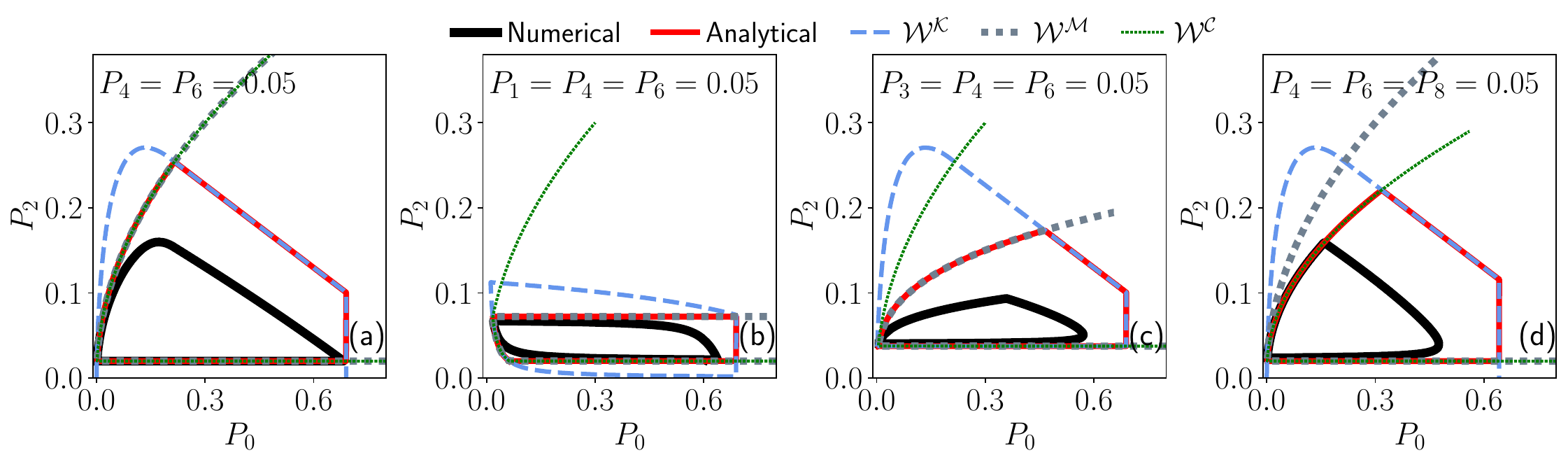}
    \caption{ Illustration how the performance of  counting-matrix withess ${\cal W}^{\cal C}$ depends on the particular set of available occupation probabilities with respect to other criteria. In all panels, we assume access to $P_0$, $P_2$, $P_4$, and $P_6$, while panels (b)--(d) include one additional probability.
    }
\label{fig:CM}
\end{figure}

Since the counting-matrix construction with half-integer indices was introduced only recently and is discussed in detail in Ref.~\cite{krishnaswamy2026observable}, we examine its relation to the other analytical criteria separately. Figure~\ref{fig:CM} illustrates how the performance of the different criteria depends on the particular set of available occupation probabilities. In all panels, we assume access to $P_0$, $P_2$, $P_4$, and $P_6$, while panels (b)--(d) include one additional probability.

For the four probabilities $\{P_0,P_2,P_4,P_6\}$ in panel (a), the largest admissible counting matrices are the two $2\times2$ matrices
\begin{equation}
\begin{pmatrix}
P_0 & 2P_2\\
2P_2 & 24P_4
\end{pmatrix},
\qquad
\begin{pmatrix}
2P_2 & 24P_4\\
24P_4 & 720P_6
\end{pmatrix},
\end{equation}
corresponding to the progressions $\{P_0,P_2,P_4\}$ and $\{P_2,P_4,P_6\}$, respectively. In this case, ${\cal W}^{\cal C}$ does not provide a criterion beyond those obtained from the majorization rules, and the corresponding boundaries coincide.

In panel (b), the additional probability $P_1$ allows one to construct, in addition, the $2\times2$ matrix
\begin{equation}
\begin{pmatrix}
P_0 & P_1\\
P_1 & 2P_2
\end{pmatrix}.
\end{equation}
The additional information improves all families of analytical criteria, but the majorization criteria ${\cal W}^{\cal M}$ benefit more strongly from it than ${\cal W}^{\cal C}$.

Panel (c) shows the corresponding situation when $P_3$, rather than $P_1$, is added. The largest counting matrices are again only $2\times2$. In addition to those already available in panel (a), one can construct, for example,
\begin{equation}
\begin{pmatrix}
P_0 & 6P_3\\
6P_3 & 720P_6
\end{pmatrix},
\qquad
\begin{pmatrix}
2P_2 & 6P_3\\
6P_3 & 24P_4
\end{pmatrix},
\end{equation}
from the progressions $\{P_0,P_3,P_6\}$ and $\{P_2,P_3,P_4\}$. Also in this case, the additional counting-matrix criteria do not outperform the majorization criteria.

A qualitatively different situation occurs in panel (d), where $P_8$ is added. The probabilities $\{P_0,P_2,P_4,P_6,P_8\}$ allow the construction of the $3\times3$ counting matrix
\begin{equation}
C^{(0,2,4)}
=
\begin{pmatrix}
P_0 & 2P_2 & 24P_4\\
2P_2 & 24P_4 & 720P_6\\
24P_4 & 720P_6 & 40320P_8
\end{pmatrix}.
\end{equation}
Here ${\cal W}^{\cal C}$ detects a larger nonclassical region than ${\cal W}^{\cal M}$. Nevertheless, the complete analytical boundary still contains substantial regions determined by ${\cal W}^{\cal K}$, showing that the counting-matrix criterion is not uniformly stronger than the other analytical criteria even when a $3\times3$ matrix is available.

These examples demonstrate that the relative strength of the analytical criteria depends strongly on the particular set of available occupation probabilities. No single analytical family is uniformly optimal. In contrast, all nonclassical regions detected by these analytical criteria are contained within those identified by the numerical optimization.

\section{Additional examples for physical models}

\subsection{Jaynes--Cummings model}

We now illustrate the numerical criterion using the resonant Jaynes--Cummings (JC) model under conditions discussed in the main text. The JC model provides a convenient benchmark, since it exhibits both classical and nonclassical regions whose extent can be directly compared for different witness 
constructions.

As already stated in the main text, we assume that initially both subsystems are in thermal equilibrium, characterized by different temperatures $T_o$ and $T_q$ for the oscillator and the two-level system (qubit), respectively.  Therefore, the initial state of the two-level system is
\begin{equation}
\rho_q
=
p_e |e\rangle\langle e|
+
(1-p_e)|g\rangle\langle g|,
\label{eq:TLS}
\end{equation}
with excited-state population given by Boltzmann distribution 
\begin{equation}
p_e=
\frac{\exp\!\left(-\frac{\hbar\omega}{k_B T_q}\right)}{
     1+\exp\!\left(-\frac{\hbar\omega}{k_B T_q}\right)}.
    \label{eq:peT}
\end{equation}
The oscillator is prepared in a thermal state
\begin{equation}
\rho_o
=
\sum_{n=0}^{\infty}
\frac{\bar n^n}{(1+\bar n)^{n+1}}
|n\rangle\langle n|,
\label{eq:rhoth}
\end{equation}
where the mean occupation number
\begin{equation}
\bar n=
\left[
\exp\!\left(\frac{\hbar\omega}{k_B T_o}\right)-1
\right]^{-1}
\end{equation} is determined by the oscillator temperature. As in the main text, the Jaynes-Cummings Hamiltonian is given by
\begin{equation}
H =
\frac{\hbar\omega}{2}\sigma_z
+\hbar\nu a^\dagger a
+\hbar g \left(a^\dagger\sigma_- + a\sigma_+\right),
\end{equation}
where we assume $\hbar=1$ and the resonant condition $\nu=\omega$. This allows a transformation to an interaction picture resulting in the interaction Hamiltonian
\begin{equation}
H_I =
g \left(a^\dagger\sigma_- + a\sigma_+\right).
\end{equation}
Observe that $\sigma_-=\ket{g}\bra{e}$ and $\sigma_+=\ket{e}\bra{g}$, and therefore $\sigma_+\sigma_-=\ket{e}\bra{e}$ and $\sigma_-\sigma_+=\ket{g}\bra{g}$. Note that $\ket{g}\bra{g}$ and $\ket{e}\bra{e}$ are idempotent and that $\sigma_\pm^n=\sigma_\pm$ for odd $n$ and 0 otherwise. This means that the unitary operator $U=e^{-iH_It}=\sum_n\frac{(-igt)^n}{n!}\left(a^\dagger\sigma_- + a\sigma_+\right)^n$ can be written in the form
\begin{equation}
    U =
    A_{gg}\ket{g}\bra{g}+A_{ee}\ket{e}\bra{e}+A_{ge}\ket{g}\bra{e}+A_{eg}\ket{e}\bra{g}\,.
\end{equation}
Evaluating a few terms of $\left(a^\dagger\sigma_- + a\sigma_+\right)^n$ is enough to see the pattern:
\begin{center}
\begin{tabular}{c|c}
    $n=0$ &  $\mathbb{I}\otimes\left(\ket{g}\bra{g}+\ket{e}\bra{e}\right)$\\
    $n=1$ & $a^\dagger\ket{g}\bra{e}+a\ket{e}\bra{g}$\\
    $n=2$ & $a^\dagger a\ket{g}\bra{g}+aa^\dagger\ket{e}\bra{e}$\\
    $n=3$ & $aa^\dagger a\ket{e}\bra{g}+a^\dagger aa^\dagger\ket{g}\bra{e}$
\end{tabular}
\end{center}
Consider the $\ket{g}\bra{e}$ terms. We have
\begin{align}
    A_{ge}&=a^\dagger\sum_{n=0}\frac{(-igt)^{2n+1}}{(2n+1)!}\left(aa^\dagger\right)^n\\
    &=a^\dagger\sum_{n=0}\frac{(-igt)^{2n+1}}{(2n+1)!}\frac{\left(\sqrt{aa^\dagger}\right)^{2n+1}}{\sqrt{aa^\dagger}}\\
    &=a^\dagger\frac{\sinh\left(-igt\sqrt{\hat n+1}\right)}{\sqrt{\hat n+1}}\\
    &=a^\dagger\frac{-i\sin\left(gt\sqrt{\hat n+1}\right)}{\sqrt{\hat n+1}}\,,
\end{align}
where we have used $\hat n=a^\dagger a$ and $[a,a^\dagger]=1$ as well as the definition of the hyperbolic sine and its imaginary argument. A similar argument allows 
\begin{align}
    A_{eg}&=a\sum_{n=0}\frac{(-igt)^{2n+1}}{(2n+1)!}\left(a^\dagger a\right)^n\\
    &=a\frac{-i\sin\left(gt\sqrt{\hat n}\right)}{\sqrt{\hat n}}\,.
\end{align}
The $\ket{g}\bra{g}$ terms evaluate to
\begin{align}
    A_{gg}&=\sum_{n=0}\frac{(-igt)^{2n}}{(2n)!}\left(a^\dagger a\right)^n\\
    &=\sum_{n=0}\frac{(-igt)^{2n}}{(2n)!}\left(\sqrt{a^\dagger a}\right)^{2n}\\
    &=\cosh\left(-igt\sqrt{\hat n}\right)\\
    &=\cos\left(gt\sqrt{\hat n}\right)\,,
\end{align}
and similarly,
\begin{equation}
    A_{ee}
    =\cos\left(gt\sqrt{\hat n+1}\right)\,,
\end{equation}

Tracing out the two-level system, the evolved oscillator state takes the form
\begin{equation}
    \rho_o(t)=p_e\left(A_{ee}\rho_oA_{ee}^\dagger+A_{ge}\rho_oA_{ge}^\dagger\right)+(1-p_e)\left(A_{gg}\rho_oA_{gg}^\dagger+A_{eg}\rho_oA_{eg}^\dagger\right)\,.
\end{equation}
Observe that
\begin{align}
    A_{ee}\rho_oA_{ee}^\dagger&=\sum_{n=0}\frac{\bar n^n}{(\bar n+1)^{n+1}}\cos^2\left(gt\sqrt{n+1})\right)\ket{n}\bra{n}\\
    A_{gg}\rho_oA_{gg}^\dagger&=\sum_{n=0}\frac{\bar n^n}{(\bar n+1)^{n+1}}\cos^2\left(gt\sqrt{n})\right)\ket{n}\bra{n}\\
    A_{eg}\rho_oA_{eg}^\dagger&=\sum_{n=1}\frac{\bar n^n}{(\bar n+1)^{n+1}}\sin^2\left(gt\sqrt{n})\right)\ket{n-1}\bra{n-1}=\sum_{n=0}\frac{\bar n^{n+1}}{(\bar n+1)^{n+2}}\sin^2\left(gt\sqrt{n+1})\right)\ket{n}\bra{n}\\
    A_{ge}\rho_oA_{ge}^\dagger&=\sum_{n=0}\frac{\bar n^n}{(\bar n+1)^{n+1}}\sin^2\left(gt\sqrt{n+1})\right)\ket{n+1}\bra{n+1}=\sum_{n=0}\frac{\bar n^{n-1}}{(\bar n+1)^n}\sin^2\left(gt\sqrt{n})\right)\ket{n}\bra{n}
\end{align}
where in the last sum of the last line we have omitted the $n=-1$ term, as the coefficient evaluates to 0. With this~\cite{marek2016deterministic,Slodicka2016deterministic}, the evolved oscillator state can be written in the form
\begin{equation}
\rho_o(\tau)
=
\sum_{n=0}^{\infty}
P_n(\tau)
|n\rangle\langle n|,
\end{equation}
where $\tau=gt$ and
\begin{align}
P_n(\tau)
&=
\frac{\bar n^n}{(1+\bar n)^{n+1}}
\Bigg\{
p_e
\left[
\cos^2\!\bigl(\tau\sqrt{n+1}\bigr)
+
\frac{1+\bar n}{\bar n}
\sin^2\!\bigl(\tau\sqrt n\bigr)
\right]
\nonumber\\
&
+
(1-p_e)
\left[
\cos^2\!\bigl(\tau\sqrt n\bigr)
+
\frac{\bar n}{1+\bar n}
\sin^2\!\bigl(\tau\sqrt{n+1}\bigr)
\right]
\Bigg\}.
\label{eq:PnJC}
\end{align}

Note that when both subsystems are initially at the same temperature, corresponding to the condition
\begin{equation}
\bar n=\frac{p_e}{1-2p_e},
\label{eq:npe}
\end{equation}
the global thermal equilibrium state is recovered and no time evolution occurs.

\begin{figure}[ht!]
    \includegraphics[width=1.0\columnwidth]{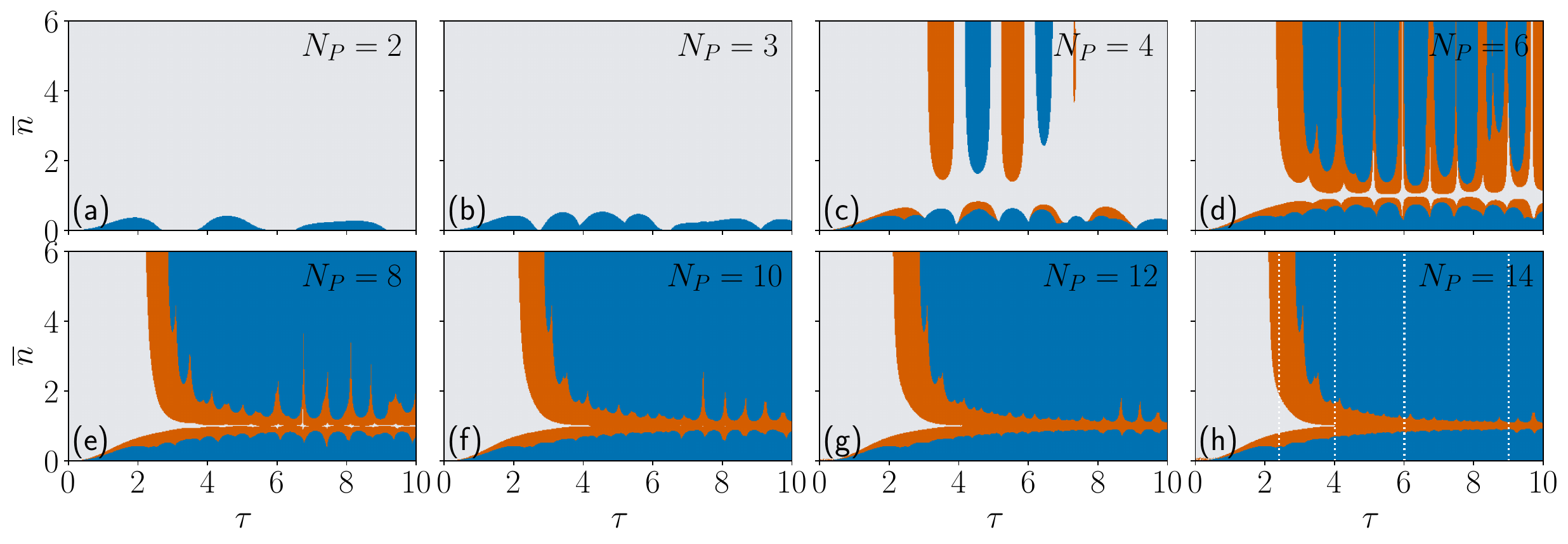}
    \caption{
    Comparison of nonclassical regions obtained using analytical and numerical criteria for the Jaynes--Cummings model. The dark blue region represents nonclassicality detected by the analytical criterion. The numerical criterion detects the same region and, for $N_P>3$, extends the nonclassical region as indicated by the orange areas. Each panel corresponds to a test performed using the first $N_P$ photon-number probabilities $P_k$. The vertical dashed lines in panel (h) correspond to the cuts analyzed in Figs.~\ref{fig:jc2} and \ref{fig:jc3}.
    }
    \label{fig:jc1}
\end{figure}

Figure~\ref{fig:jc1} presents the nonclassicality maps for the case $p_e=1/3$, corresponding to $k_BT/\hbar\omega=1.44$, which complements the examples shown in the main text. The dark blue regions denote states identified as nonclassical by the analytical criterion ${\cal W}^{\cal K}$, while the orange regions correspond to additional nonclassicality detected only by the numerical optimization. For $N_P=3$ both approaches give identical results, reflecting the fact that the available analytical witnesses are already optimal in this case. Increasing the number of probabilities immediately enlarges the nonclassical region detected by the numerical method, with the first noticeable improvement appearing already for $N_P=4$.
\begin{figure}[ht!]
    \includegraphics[width=1.0\columnwidth]{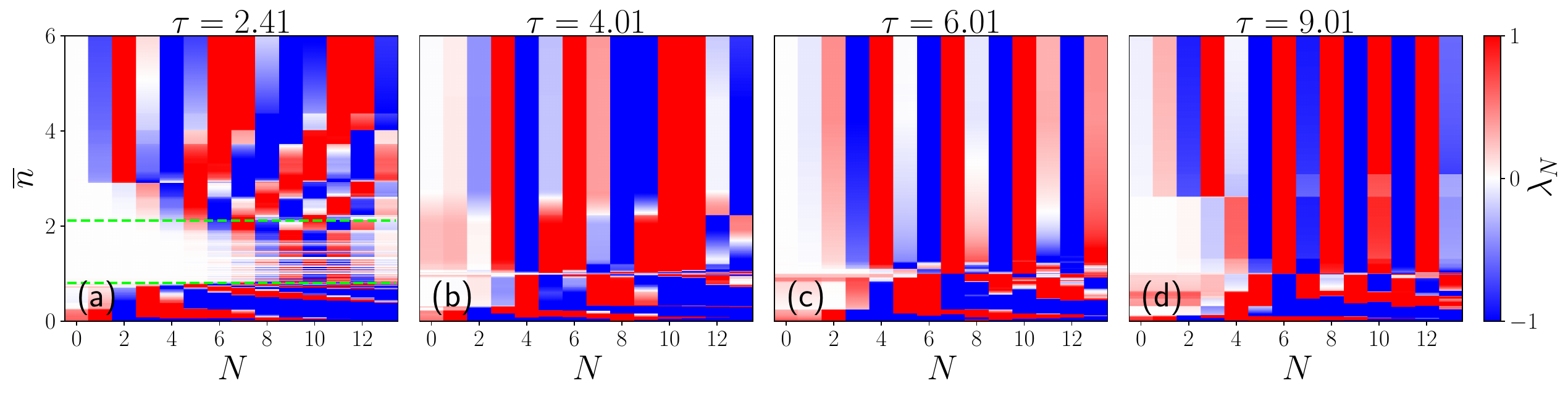}
    \caption{
    Evolution of the parameters $\bm{\lambda}$ of the nonclassicality witnesses obtained from the numerical criterion. Individual panels correspond to the cuts at different values of $\tau$ indicated in Fig.~\ref{fig:jc1}. The horizontal dashed lines in panel (a) mark the region that is not identified as nonclassical by the criterion.
    }
    \label{fig:jc2}
\end{figure}

Two features are particularly relevant from the experimental point of view. First, the additional nonclassical regions appear predominantly at shorter interaction times $\tau$, which are generally easier to access experimentally because decoherence effects are weaker. Second, the numerical criterion substantially reduces the region classified as classical around the thermal-equilibrium condition Eq.~\eqref{eq:npe},
where the initial state of the qubit and oscillator corresponds to the global thermal equilibrium. At this point no time evolution occurs and the state therefore remains classical. Nevertheless, the numerical optimization reveals sizeable nonclassical regions in its immediate vicinity that are not detected by the analytical witnesses.
\begin{figure}[ht!]
    \includegraphics[width=0.6\columnwidth]{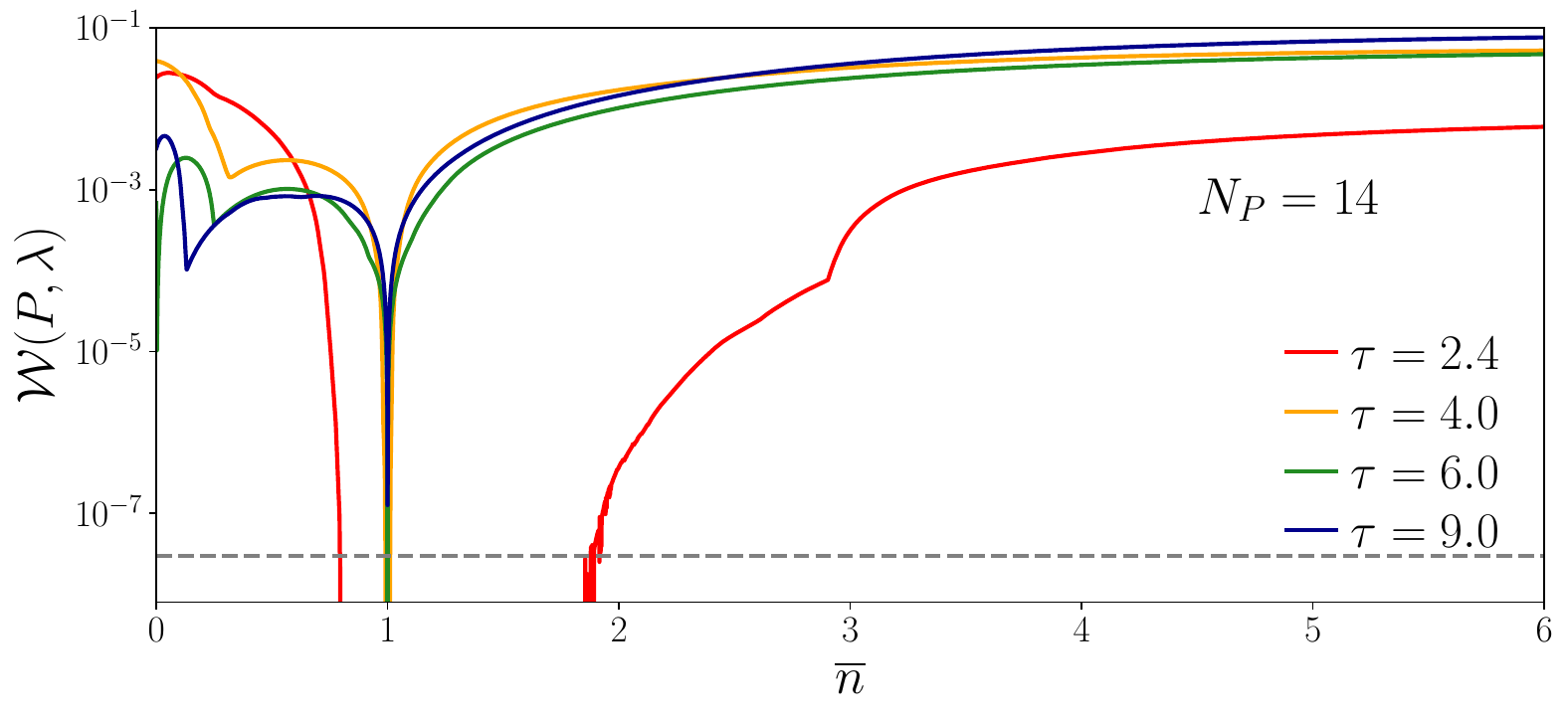}
    \caption{
    Value of the witness function
    ${\mathcal W}(\boldsymbol\lambda;\mathbf P)
=\boldsymbol\lambda\cdot\mathbf P
-\sup_{\mu\geq0}
\boldsymbol\lambda\cdot\mathbf p(\mu)$
    for selected cuts through the parameter space. The figure illustrates the stability of the numerical implementation in different regimes. Of particular interest is the red curve ($\tau=2.4$), for which the nonclassical region at larger values of $\bar n$ is not detected by the analytical criterion. The horizontal dashed line indicates the threshold used for identifying nonclassicality.
    }
    \label{fig:jc3}
\end{figure}

The behavior of the numerical optimization itself is illustrated in Figs.~\ref{fig:jc2} and \ref{fig:jc3}, which correspond to the cuts indicated in Fig.~\ref{fig:jc1}(h) by vertical dashed lines. Figure~\ref{fig:jc2} shows the optimized coefficients $\bm{\lambda}$ defining the numerical witness. Although several coefficients approach the limiting values $\pm1$ inside the nonclassical region, no simple or universal pattern emerges that could be used to infer improved analytical witnesses. Instead, the optimal coefficients vary across the parameter space, illustrating that the numerical optimization exploits the available occupation probabilities in a highly nontrivial manner.
\begin{figure}[ht!]
    \includegraphics[width=0.9\columnwidth]{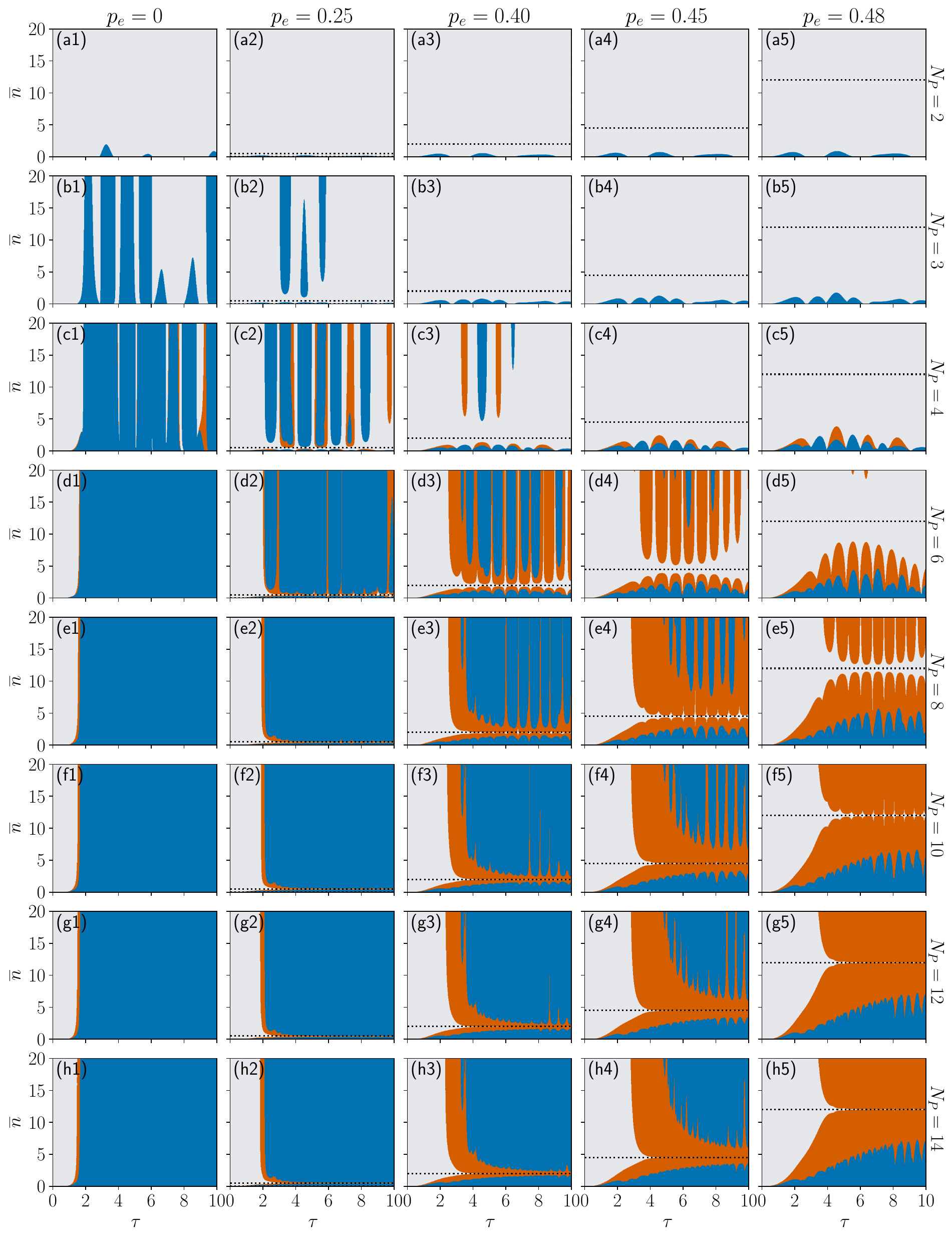}
    \caption{
    Comparison of nonclassical regions obtained using analytical ${\cal W}^{\cal K}$ and numerical criteria for the Jaynes--Cummings model for different values of the initial excited-state population $p_e$ and different numbers of probabilities $N_P$ used in the criteria. The dark blue region represents nonclassicality detected by the analytical criterion, while the orange regions correspond to additional nonclassicality detected only by the numerical criterion. The horizontal dotted line indicates the thermal-equilibrium condition given by Eq.~\eqref{eq:npe}.
    }
    \label{fig:jcBig}
\end{figure}

The corresponding optimized witness values are shown in Fig.~\ref{fig:jc3}. The witness remains smooth throughout the parameter range, demonstrating the numerical stability of the semi-infinite linear-program algorithm. The pronounced minimum near the thermal-equilibrium condition Eq.~\eqref{eq:npe} reflects the gradual "disappearance of nonclassicality" as the system approaches thermal equilibrium. The apparent closing of this minimum for longer interaction times Fig.~\ref{fig:jc1}(h) is a resolution effect caused by the finite discretization of $\bar n$ rather than a genuine disappearance of the classical region as it is clear from the sharp minima dips in Fig.~\ref{fig:jc3} for $\tau\geq 6$.

\begin{figure}[ht!]
    \includegraphics[width=1.0\columnwidth]{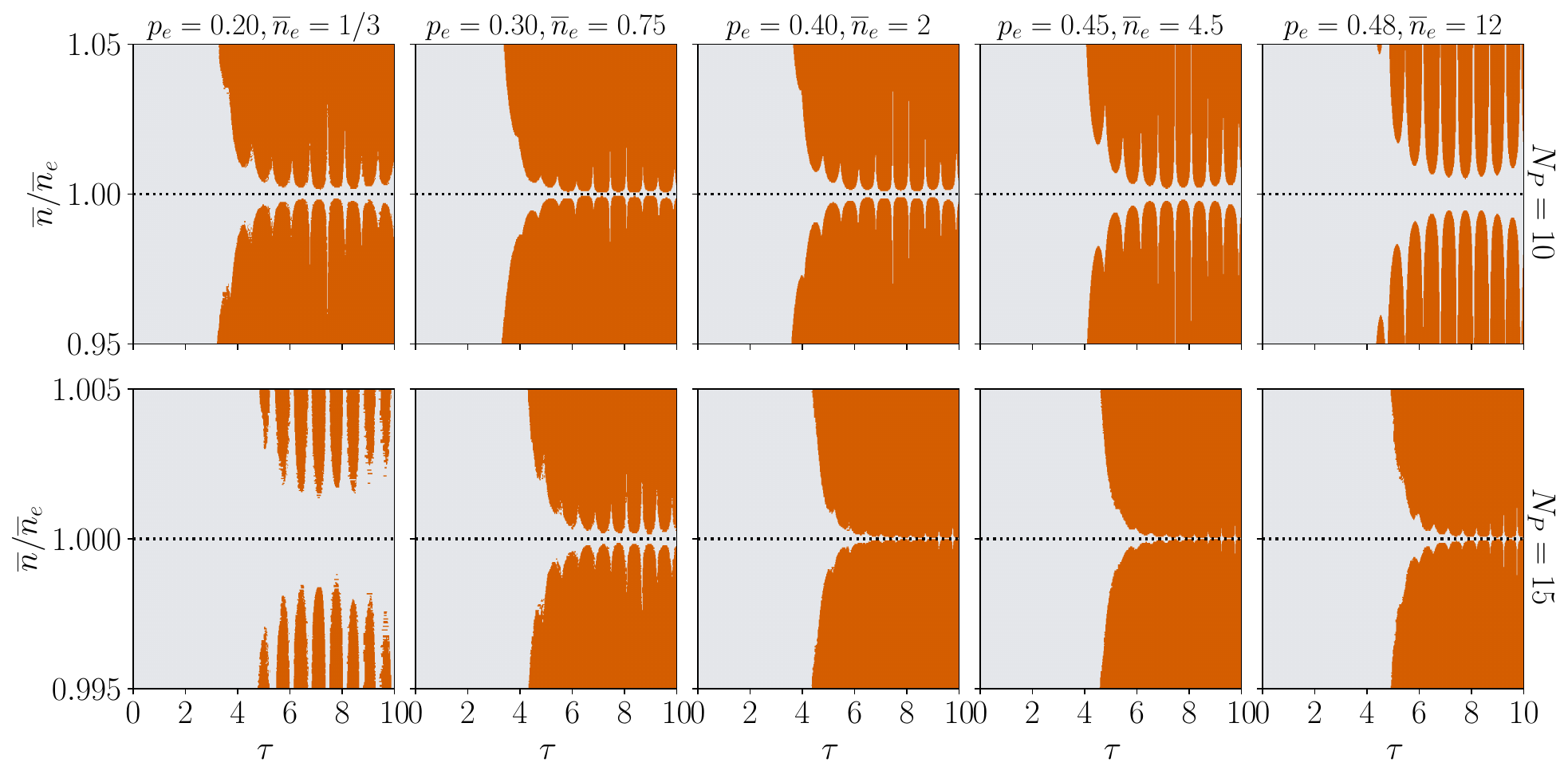}
    \caption{
    Detail of the vicinity of the thermal-equilibrium condition (horizontal dotted line) given by Eq.~\eqref{eq:npe} for several values of $p_e$. The first row corresponds to $N_P=10$, while the second row shows results for $N_P=15$.
    }
    \label{fig:jcFocus}
\end{figure}

Finally, Figs.~\ref{fig:jcBig} and \ref{fig:jcFocus} demonstrate that the conclusions drawn above remain valid throughout the parameter space. Figure~\ref{fig:jcBig} presents nonclassicality maps for several values of the initial excited-state population $p_e$, while Fig.~\ref{fig:jcFocus} shows the detail of the vicinity of the thermal-equilibrium condition. As $p_e$ approaches $1/2$, the detection of nonclassical region becomes harder. Increasing $N_P$ from 10 to 15 noticeably enlarges the detected nonclassical region, indicating that higher occupation probabilities continue to contain useful information about nonclassicality even close to thermal equilibrium.

Before turning to this regime, let us briefly discuss several technical aspects of the numerical criterion. Figure~\ref{fig:jc2} shows the evolution of the optimal parameters $\bm{\lambda}$ entering the numerically constructed nonclassicality witnesses. Within the nonclassical region, several components of $\bm{\lambda}$ often approach values close to $\pm 1$. However, we have not identified a simple or systematic pattern that could be used to construct explicit analytical witnesses from these numerical solutions.

Figure~\ref{fig:jc3} shows the value of the witness function ${\cal W}(\bm{\lambda};\bm{P}_{\mathcal{I}})$, Eq.~\eqref{eq:SM_witness_num} used in the numerical criterion for $p_e=1/3$, different interaction times $\tau$ and initial occupations $\bar n$. The figure illustrates the numerical stability of the optimization procedure across a broad range of parameters.

A pronounced dip is visible near the equilibrium condition given by Eq.~\eqref{eq:npe}, where the value of the witness falls below the threshold used to identify nonclassicality. This region becomes narrower with increasing interaction time $\tau$. Although it appears to close completely for sufficiently long times, a more detailed analysis reveals that this is a consequence of the finite resolution in $\bar n$ rather than a genuine disappearance of the classical region.

Figure~\ref{fig:jcBig} presents an extended set of nonclassicality maps analogous to those shown in main text. The results demonstrate that the role of the numerical criterion becomes increasingly important as $p_e$ approaches $1/2$. In particular, a larger number of probabilities $N_P$ is required to resolve the nonclassicality in the vicinity of the equilibrium condition~\eqref{eq:npe}.
A detailed view of this region for different values of $p_e$ is shown in Fig.~\ref{fig:jcFocus}.

These results highlight the importance of considering a sufficiently large number of occupation probabilities when analyzing nonclassicality close to the thermal-equilibrium condition. At exact equilibrium, corresponding to $\bar n=\bar n_e$, the state does not evolve and therefore remains classical. Nevertheless, the numerical analysis reveals nonclassical regions in the vicinity of this point. Resolving these regions requires access to larger values of $N_P$, for which the numerical criterion provides a substantial advantage over the available analytical witnesses.

\FloatBarrier
\subsection{Boson-added coherent states: Probabilistic boson addition}

To complement the nonlinear and thermally initialized Jaynes--Cummings
dynamics, we consider a much simpler but qualitatively different scenario: probabilistic boson addition to a coherent state. Here the initial state is pure and at zero temperature, and nonclassicality is generated by a single conditional quantum operation rather than by coherent nonlinear dynamics.

As a reference, a single boson added deterministically to a coherent state produces a nonclassical state with analytically known occupation
probabilities, which is readily identified by the available analytical
witnesses~\cite{innocenti_nonclassicality_2022}. The more challenging case arises when the addition succeeds only with probability $p$, while with probability $1-p$ the original coherent state is left unchanged. The resulting state is therefore
\begin{equation}
\rho
=
p\,
\frac{\hat a^\dagger |\alpha\rangle\langle\alpha|\hat a}
{\langle\alpha|\hat a\hat a^\dagger|\alpha\rangle}
+
(1-p)\,
|\alpha\rangle\langle\alpha|,
\end{equation}
which continuously interpolates between a coherent state ($p=0$) and the single-boson-added coherent state ($p=1$). The corresponding occupation probabilities are
\begin{equation}
P_k
=
\frac{e^{-\mu}\mu^k}{k!}
\left[
(1-p)
+
p\,
\frac{k^2}{\mu(\mu+1)}
\right],
\label{eq:pba_probabilities}
\end{equation}

where $\mu=|\alpha|^2$ denotes the average occupation of the initial coherent state.
\begin{figure}[ht!]
	\includegraphics[width=1.0\columnwidth]{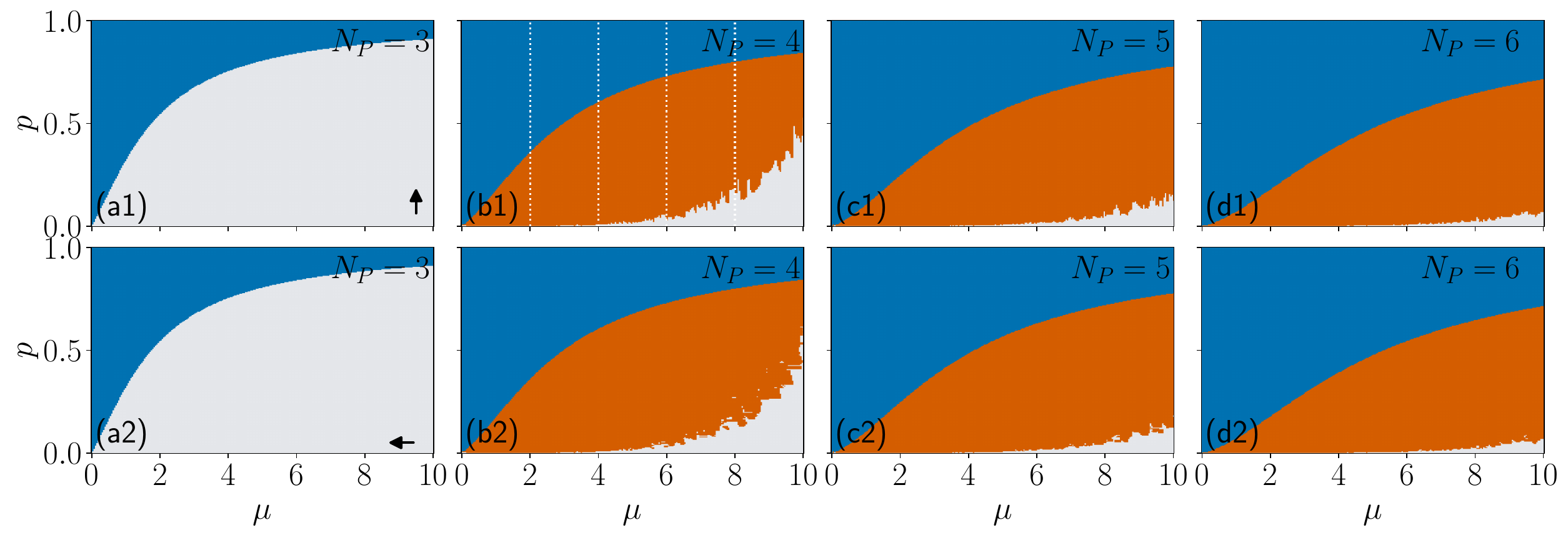}\caption{
Comparison of nonclassical regions obtained using analytical and numerical criteria for a model with probabilistic single-boson addition. The dark blue region represents the nonclassicality detected by the analytical criterion. The numerical criterion detects the same region and, for $N_P>3$, further extends the nonclassical region, as indicated in orange. Each panel corresponds to a test performed using the first $N_P$ photon-number probabilities $P_k$. The first row shows a strategy in which the solution for $p_i$ is propagated to $p_{i+1}$ (i.e., used as an initial guess before solving the corresponding linear program), while the second row shows the reverse strategy, where the solution for $\mu_{i+1}$ is propagated to $\mu_i$.~\label{fig:pb1}}
\end{figure}

Figure~\ref{fig:pb1} compares the analytical witnesses with the numerical optimization for this model. As in the Jaynes--Cummings example, the dark blue regions correspond to states identified by the analytical criteria, whereas the orange regions denote additional nonclassicality detected only by the numerical approach. The improvement again becomes visible once more than three occupation probabilities are included in the witness construction.
\begin{figure}[ht!]
	\includegraphics[width=1.0\columnwidth]{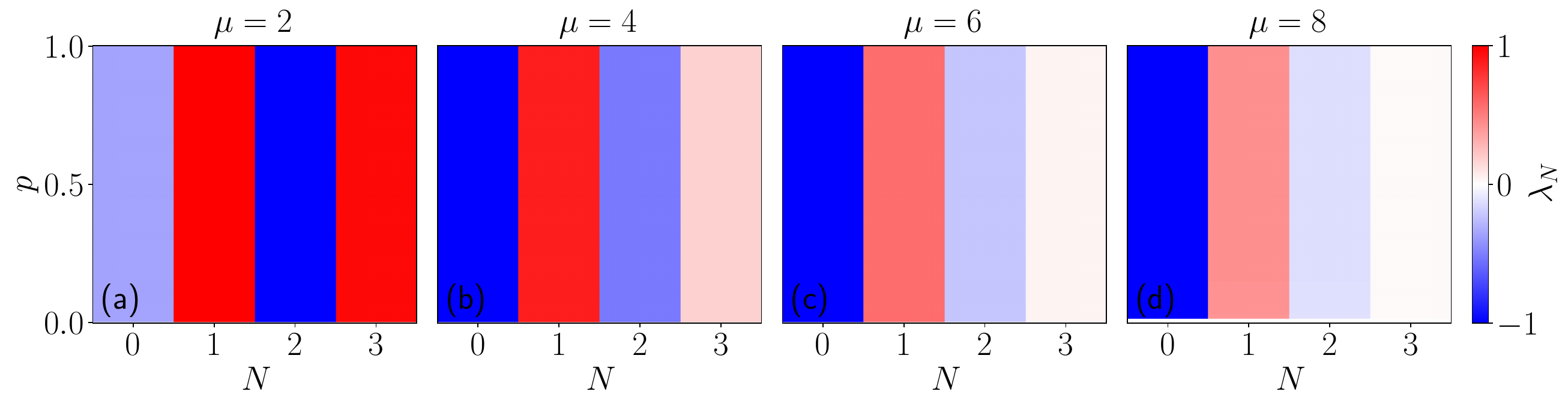}\caption{
Evolution of the parameters $\bm{\lambda}$ of the nonclassicality witnesses obtained from the numerical criterion. The computation is performed sequentially in the direction of increasing $p$, where the solution obtained for a smaller value of $p$ is first tested as an initial candidate before solving the corresponding linear program. Individual panels correspond to cuts at different values of $\mu$, as shown in Fig.~\ref{fig:pb1}(b1). Note that once a given combination $\bm{\lambda}(p)$ is accepted, it satisfies the criterion for all $p$ from the critical value $p_c$ up to $p=1$.
\label{fig:pb2}}
\end{figure}

In contrast to the Jaynes--Cummings model, the parameter space is now naturally divided into regions with weak and strong boson-addition probability. For small values of $p$, the state remains close to the original coherent state and its nonclassicality becomes increasingly difficult to certify. The numerical optimization substantially enlarges the detected region, especially for larger values of the coherent-state occupation $\mu$, where the analytical witnesses rapidly lose sensitivity.
\begin{figure}
	\includegraphics[width=0.8\columnwidth]{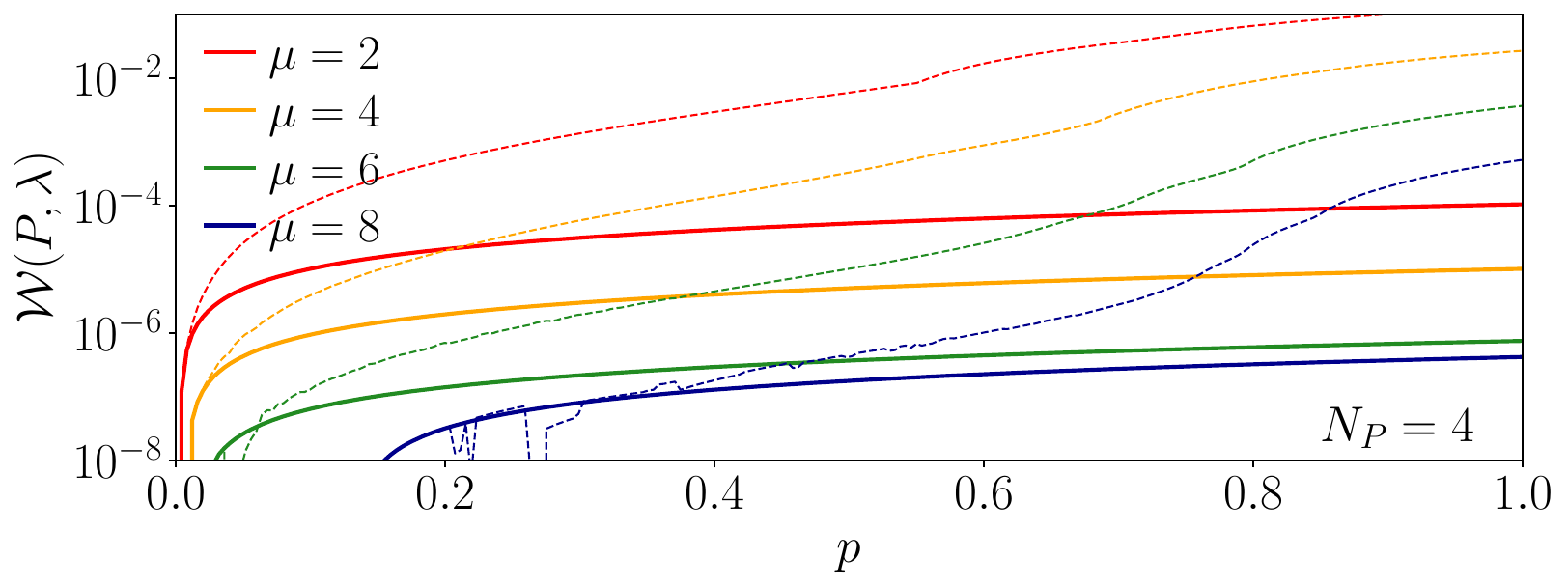}\caption{
Comparison of the witness function (quantum--classical gap) evaluated using two protocols. The solid lines correspond to a sequential protocol, in which the solution obtained at $p_i$ is propagated and tested at $p_{i+1}$ before solving the corresponding linear program. The dashed lines represent an independent protocol, where each point is treated separately. The independent protocol can yield parameters $\bm{\lambda}$ that produce a witness value up to three orders of magnitude larger than that obtained with the sequential method. However, using the solution $\bm{\lambda}$ at the critical point $p_c$ yields a finite gap for all $p>p_c$, which significantly accelerates the computation in the dependent protocol.   
 ~\label{fig:pb3}}
\end{figure}

The two rows of Fig.~\ref{fig:pb1} compare two equivalent strategies for constructing the nonclassicality maps. In the first row, the optimization proceeds sequentially from smaller to larger values of $p$, using the witness obtained at one point as the initial candidate for the next one. In the second row, the optimization is performed in the direction of decreasing $\mu$. The underlying idea is that once a witness, i.e., a set of coefficients $\bm{\lambda}$, certifies a state as nonclassical, it often remains valid for states with larger values of $p$ or smaller values of $\mu$, respectively. This substantially reduces the number of linear programs that need to be solved.

Both strategies produce essentially identical nonclassicality maps. Small differences are visible only close to the boundary between the classical and nonclassical regions, where numerical precision and the safety tolerances used in the optimization become important. As demonstrated in our example codes~\cite{code_github}, these residual differences can be further reduced by systematically revisiting the boundary points. Moreover, since it is sufficient to find \emph{any} witness that certifies nonclassicality, combining several propagation strategies or using different initial candidates provides an effective way of refining and smoothing the numerical boundary.

Figure~\ref{fig:pb2} provides further insight into the behavior of the optimized witnesses by showing the evolution of the coefficients $\bm{\lambda}$ along the cuts indicated in Fig.~\ref{fig:pb1}(b1). In particular, once a witness first certifies nonclassicality at the critical probability $p_c$, it frequently continues to satisfy the criterion up to $p=1$. This explains why propagating previously obtained witnesses is so effective: over a large part of the parameter space, the optimal witness changes only at a relatively small number of points. Consequently, many linear programs can be avoided simply by testing whether a previously determined witness already certifies the new state.

This behavior is quantified in Fig.~\ref{fig:pb3}, where the witness value
${\cal W}(\bm{\lambda},\bm{P})$
is compared for the sequential (solid lines) and independent (dashed lines) optimization protocols. Solving every point independently often produces witness functions with considerably larger positive values, even by orders of magnitude, than those obtained by propagating previously found solutions. Nevertheless, the sequential approach already yields a finite witness throughout the entire nonclassical region once the critical point has been identified. Since the computational cost is dominated by solving the linear programs rather than by evaluating the witness itself, this observation provides a substantial practical speed-up when constructing dense nonclassicality maps.

Overall, this example illustrates that the numerical witnesses are particularly advantageous for weakly nonclassical states that are close to the classical boundary. At the same time, it demonstrates that the numerical algorithm can exploit the smooth evolution of the optimal witnesses across parameter space, making the computation of large phase diagrams considerably more efficient.

%